\documentclass[twocolumn]{aastex63}

\newcommand\SFRD{\mathcal{SFRD}}

\usepackage{comment}

\received{July 27, 2021}
\revised{January 31, 2022}
\accepted{February, 2022}
\submitjournal{ApJ}

\shorttitle{Star formation in GOODS-N}
\shortauthors{Enia et al.}

\begin{document}

\title{A new estimate of the cosmic star formation density from a radio-selected sample, \\ and the contribution of $H$-dark galaxies at $z \geq 3$}

\correspondingauthor{Andrea Enia, Margherita Talia}
\email{andrea.enia@unibo.it, margherita.talia2@unibo.it}

\author[0000-0002-0200-2857]{Andrea Enia}
\author[0000-0003-4352-2063]{Margherita Talia}
\author[0000-0002-3592-2131]{Francesca Pozzi}
\affiliation{University of Bologna - Department of Physics and Astronomy “Augusto Righi” (DIFA), Via Gobetti 93/2, I-40129, Bologna, Italy}
\affiliation{INAF - Osservatorio di Astrofisica e Scienza dello Spazio, Via Gobetti 93/3, I-40129, Bologna, Italy}

\author[0000-0002-4409-5633]{Andrea Cimatti}
\affiliation{University of Bologna - Department of Physics and Astronomy “Augusto Righi” (DIFA), Via Gobetti 93/2, I-40129, Bologna, Italy}
\affiliation{INAF - Osservatorio Astrofisico di Arcetri, Largo E. Fermi 5, I-50125, Firenze, Italy}

\author[0000-0001-8706-2252]{Ivan Delvecchio}
\affiliation{INAF—Osservatorio Astronomico di Brera, via Brera 28, I-20121, Milano, Italy}

\author[0000-0002-2318-301X]{Gianni Zamorani}
\affiliation{INAF - Osservatorio di Astrofisica e Scienza dello Spazio, Via Gobetti 93/3, I-40129, Bologna, Italy}

\author[0000-0002-9948-0897]{Quirino D'Amato}
\affiliation{INAF/IRA, Istituto di Radioastronomia, Via Piero Gobetti 101, 40129 Bologna, Italy}
\affiliation{University of Bologna - Department of Physics and Astronomy “Augusto Righi” (DIFA), Via Gobetti 93/2, I-40129, Bologna, Italy}

\author[0000-0003-0492-4924]{Laura Bisigello}
\affiliation{INAF - Osservatorio di Astrofisica e Scienza dello Spazio, Via Gobetti 93/3, I-40129, Bologna, Italy}

\author[0000-0002-5836-4056]{Carlotta Gruppioni}
\affiliation{INAF - Osservatorio di Astrofisica e Scienza dello Spazio, Via Gobetti 93/3, I-40129, Bologna, Italy}
\author[0000-0002-9415-2296]{Giulia Rodighiero}
\affiliation{University of Padova - Department of Physics and Astronomy “Galileo Galilei”, Vicolo dell'Osservatorio 3, I-35141, Padova, Italy}
\author[0000-0002-6175-0871]{Francesco Calura}
\affiliation{INAF - Osservatorio di Astrofisica e Scienza dello Spazio, Via Gobetti 93/3, I-40129, Bologna, Italy}

\author[0000-0003-1246-6492]{Daniele Dallacasa}
\affiliation{University of Bologna - Department of Physics and Astronomy “Augusto Righi” (DIFA), Via Gobetti 93/2, I-40129, Bologna, Italy}

\author[0000-0002-1847-4496]{Marika Giulietti}
\affiliation{SISSA, Via Bonomea 265, I-34136 Trieste, Italy}

\author[0000-0003-3419-538X]{Luigi Barchiesi}
\affiliation{University of Bologna - Department of Physics and Astronomy “Augusto Righi” (DIFA), Via Gobetti 93/2, I-40129, Bologna, Italy}

\author[0000-0002-6444-8547]{Meriem Behiri}
\affiliation{SISSA, Via Bonomea 265, I-34136 Trieste, Italy}

\author[0000-0002-9948-3916]{Michael Romano}
\affiliation{University of Padova - Department of Physics and Astronomy “Galileo Galilei”, Vicolo dell'Osservatorio 3, I-35141, Padova, Italy}
\affiliation{INAF - Osservatorio Astronomico di Padova, Vicolo dell'Osservatorio 5, I-35122, Padova, Italy}

\begin{abstract}
The Star Formation Rate Density (SFRD) history of the Universe is well constrained up to redshift $z \sim 2$. At earlier cosmic epochs, the picture has been largely inferred from UV-selected galaxies (e.g. Lyman-break galaxies, LBGs). However, LBGs' inferred SFRs strongly depend on the assumed dust extinction correction, which is not well-constrained at high-$z$, while observations in the radio domain are not affected by this issue. In this work we measure the SFRD from a 1.4 GHz-selected sample of $\sim$600 galaxies in the GOODS-N field up to redshift $\sim 3.5$. We take into account the contribution of Active Galactic Nuclei from the Infrared-Radio correlation. We measure the radio luminosity function, fitted with a modified Schechter function, and derive the SFRD. The cosmic SFRD shows a rise up to $z \sim 2$ and then an almost flat plateau up to $z \sim 3.5$. Our SFRD is in agreement with the ones from other FIR/radio surveys and a factor 2 higher than those from LBG samples. We also estimate that galaxies lacking a counterpart in the HST/WFC3 H-band ($H$-dark) make up $\sim 25\%$ of the $\phi$-integrated SFRD relative to the full sample at z $\sim 3.2$, and up to $58\%$ relative to LBG samples. 
\end{abstract}


\section{Introduction} \label{sec:intro}

The cosmic history of star formation in the Universe is one of the main topics in the field of galaxy formation and evolution. The amount of stars formed per year in a unit of cosmological volume (i.e. star formation rate density, SFRD) is now well understood up to redshift $3$, when the Universe was no more than $3$ Gyrs old, thanks to a great amount of multiwavelength studies gathered in the past $30$ years (\citealp[see][for an exhaustive review]{2014ARA&A..52..415M}; \citealp[more recently][]{2018ApJ...855..105O}).

Looking backwards in time, the SFRD increases by a factor $\sim 8$ from the present day, reaching a peak at redshift $z \sim 2$, an epoch known as ``cosmic noon''.  A similar evolution is observed for the black hole accretion rate density, a proxy for galactic nuclear activity, peaking at similar redshift \citep{1998MNRAS.293L..49B, 2014MNRAS.439.2736D, 2014ARA&A..52..415M}. The picture becomes less clear at higher redshifts \citep{2014ApJ...796...95C, 2019ApJ...877...45M}. Several studies show a steep decline in the SFRD \citep{2015ApJ...803...34B, 2016MNRAS.459.3812M, 2018ApJ...854...73I}, though it could be a consequence of sample selection, the vast majority being UV-selected sources such as Lyman-break galaxies (LBGs). Studies performed at longer wavelengths (i.e. radio or sub-mm bands) show a flatter SFRD at $z > 3$ (\citealp[][]{2013MNRAS.432...23G, 2016MNRAS.461.1100R, 2017A&A...602A...5N, 2020A&A...643A...8G}; \citealp[but see also][]{2022MNRAS.509.4291M}) and a full understanding of the intrinsic evolution of SFRD in the young Universe is still uncertain.

There are various possible ways to explain this tension. For example, results based on UV-selected sources rely on the adopted dust extinction correction \citep[e.g.][]{2014ApJ...796...95C} to infer the bolometric star-formation rate (SFR), which is not well constrained at high-$z$. Since the first detections with the Submillimetre Common-User Bolometer Array \citep[SCUBA,][]{1998Natur.394..241H}, several Far-Infrared (FIR) to sub-mm surveys \citep[for an exhaustive review see][]{2014ApJ...796...95C} revealed the existence of a population of dusty star forming galaxies (DSFGs), rare in the local Universe but common at high-$z$, with extreme SFRs (up to $10^3 - 10^4$ M$_{\odot}$ yr$^{-1}$) whose dust-obscured component is directly sampled with FIR observations. However, results based on these observations are limited by the achievable sensitivity and large beam of single dish FIR/sub-mm telescopes (\citealp[e.g. Herschel,][]{2011A&A...532A..90L}; \citealp[or SCUBA,][]{1999MNRAS.303..659H}), which make it hard to reach $z > 3$ and properly identify counterparts for the most distant objects. Moreover, higher resolution surveys performed with interferometers like NOEMA and ALMA in the mm regime currently cover a too small area to map sufficiently large volumes \citep[e.g.][]{walter2016,dunlop2017,franco2018,hatsukade2018,decarli2019,gonzalezlopez2020,2020A&A...643A...1L,2020A&A...643A...2B,faisst2020,casey2021}.

Radio-selected surveys are a way to overcome some of those issues. The high resolving power reached by interferometers such as the {\it Karl G. Jansky} Very Large Array (JVLA or VLA) allows resolutions high enough to facilitate the counterpart identification with respect to FIR/sub-mm single-dish observations, overcoming the beam confusion. Once corrected for possible Active Galactic Nuclei (AGN) contamination, the radio frequencies directly sample star formation processes without being affected by dust extinction, as a consequence of the nature of non-thermal radio emission, caused by electrons accelerated in the remnants of supernova explosions of massive stars, and free-free continuum emission coming from \ion{H}{2} regions. Results obtained with radio-selected samples show a shallower evolution of the SFRD at $z > 2$ \citep{2017A&A...602A...5N} than inferred from LBGs, similar to what derived from FIR surveys. 

Moreover, LBGs are not the only star forming population at $z > 2$. It is becoming more and more evident how the information coming from rest-frame UV samples at high-redshift likely misses a non negligible amount of star formation, occurring in galaxies undetected in optical/UV bands. Previous studies indicate a contribution of these galaxies to the cosmic SFRD equal to $10\%$ of that by LBGs at $z > 3$, rising up to $\sim 25-40\%$ at $z > 4.5$ \citep{2019Natur.572..211W, 2019ApJ...884..154W, 2020A&A...643A...8G, 2021ApJ...909...23T}. These objects, usually referred to as HST-, UV-, or OIR-dark galaxies, have been selected in different ways:  serendipitously in deep CO line scan surveys \citep{2019ApJ...884..154W}, in SCUBA surveys \citep{2020ApJ...896L..21R}, via their extreme NIR color  \citep{2016ApJ...816...84W, 2019Natur.572..211W}, in continuum FIR \citep{2018A&A...620A.152F, 2019ApJ...878...73Y, 2020A&A...643A...8G,manning2021} or radio \citep{2021ApJ...909...23T} surveys. A complete census on their physical properties, their contribution to the star formation history of the Universe, as well as a proper understanding of the overlap between the different selections, is still under debate.

In this work, we make use of deep 1.4 GHz JVLA observations in the Great Observatories Origins Deep Survey field in the Northern hemisphere (GOODS-N) to measure the SFRD evolution up to $z \sim 3.5$. We also identify a sample of high-redshift galaxies undetected in the HST/WFC3 $H$-band (hereafter $H$-dark) and measure their contribution to the total cosmic SFRD.

The paper is organized as follows. In Section \ref{sec:Data} we present the multi-wavelength data set considered in this work, with particular focus on the radio 1.4 GHz observations constituting the starting point of our sample. In Section \ref{sec:Sample} we describe our sample. In Section \ref{sec:lumfunc} we present the radio luminosity function, and in Section \ref{sec:SFRDhist} we derive the SFRD and assess the contribution of the $H$-dark galaxies at $z \sim 3.5$.

Throughout this work, we give magnitudes in the AB photometric system, adopt a flat $\Lambda$-CDM cosmology with H$_0=67.8\pm0.9$ km s$^{-1}$ Mpc$^{-1}$ and $\Omega_{\rm M}=0.308\pm0.012$ \citep{2016A&A...594A..13P}, and assume a \citet{2003PASP..115..763C} initial mass function (IMF).

\section{Data}\label{sec:Data}
In this work we focus on the radio-selected sample in the GOODS-N field, and exploit the great wealth of multiwavelength ancillary data available (from X-ray to radio) and their excellent depth. The GOODS-N field \citep{2003mglh.conf..324D, 2004ApJ...600L..93G} is a well studied portion of the Northern sky, overlapping with the Hubble Deep Field North \citep[HDF-N,][]{1996AJ....112.1335W}, centered approximately at R.A. $=$ 12h36m55s, $\delta = $ 62d14m18s and covering an area of $\sim 171$ arcmin$^2$. 

\subsection{Radio data}
Our starting sample is the 1.4 GHz radio catalog presented in \citet[][hereafter O18]{2018ApJS..235...34O}. We refer to O18 for a complete description of the data analysis and multi-band association procedure. Here we summarize the points relevant for the present study.

The map reaches a root-mean-square noise of 2.2 $\mu$Jy/beam at the phase center, making it one of the deepest available in the radio, at a resolution of 1.6$\arcsec$. The catalog is extracted from different realizations of the map at varying resolution (1.6$\arcsec$, 2.0$\arcsec$, 3.0$\arcsec$, 6.0$\arcsec$ and 12.0$\arcsec$), detecting a total of 795 discrete sources down to 5$\sigma_{\rm rms}$ (with the $\sigma_{\rm rms}$ changing with the resolution).

\begin{figure}
\centering
\includegraphics[width=\columnwidth]{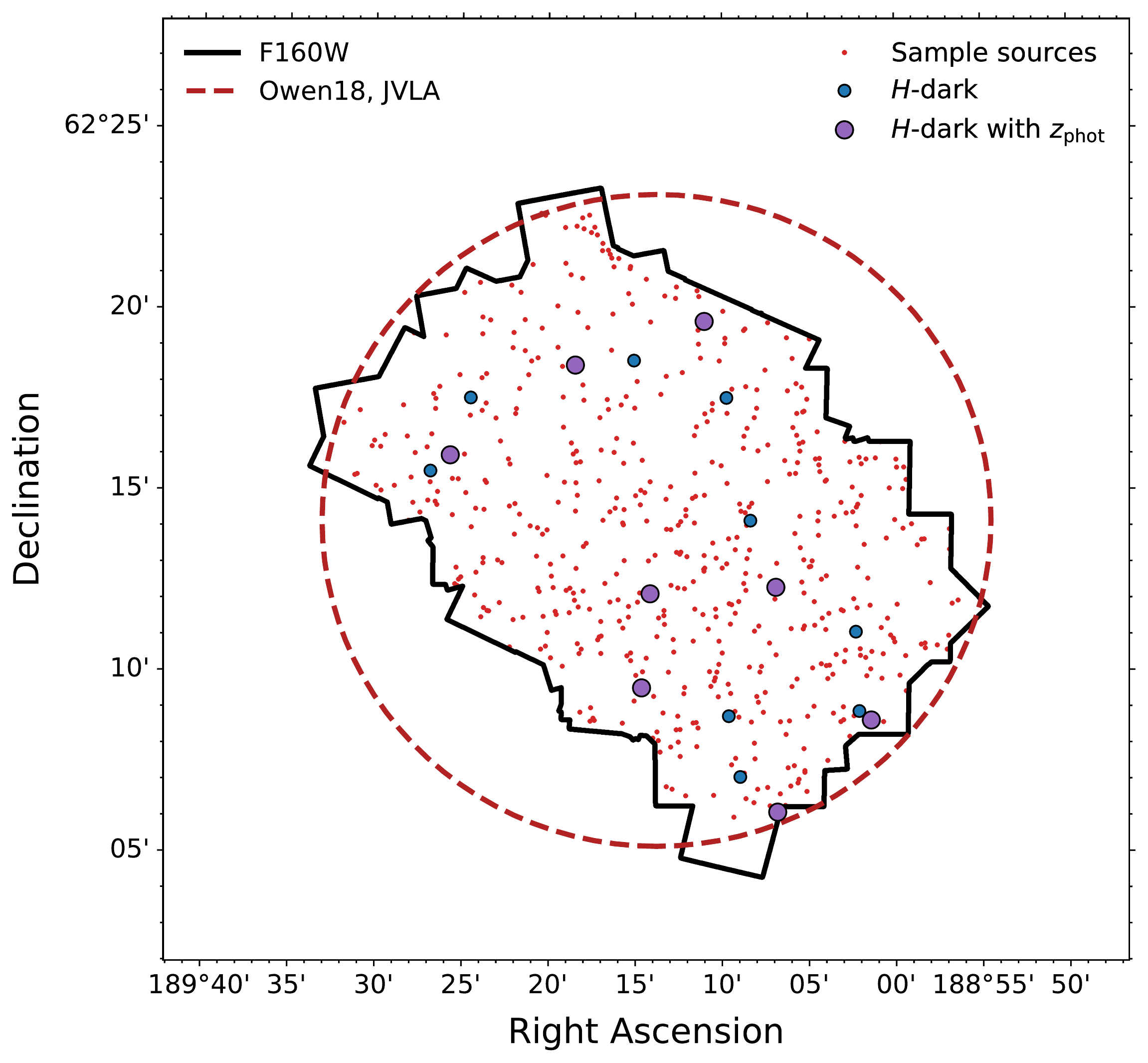}
\caption{The sample sky coverage. Black continuous line is the HST/WFC3 F160W footprint, while the red dashed circle mark the 1.4 GHz VLA observations in \citet{2018ApJS..235...34O}. Red dots are the radio sources studied in this paper, the ones identified as potential $H$-dark are colored in blue, the purple dots are the ones with photometric redshifts (see Sec.\,\ref{subsec:Hdarkweight} for a note on the $H$-dark definition).}
\label{fig:coverage}
\end{figure}

The radio observations upon which O18 builds its catalog cover an area of $\sim 9$ arcmin radius. However, given the necessity of multiwavelength data, we limited our sample to the footprints of NIR surveys, thus reducing the area of interest for our analysis to 171 arcmin$^2$ (Fig.\,\ref{fig:coverage}).

In O18, the counterpart identification is based on the deep $K_s$ catalog by \citet[][$5\sigma$ depth of 24.45]{2010ApJS..187..251W}, following a criterion based on distance reported in Eq.\,1 of O18. In the present work, we improve the association process, using a more reliable likelihood procedure instead of a nearest neighbor method, and looking for counterparts in a deeper $H$-band map (see Sec\,\ref{subsec:cpart}), down to a $5\sigma$ depth of 28.2 mag in the innermost part.

Moreover, we use two additional sets of data in the radio domain, in order to measure the radio spectral index $\alpha$ for sources with multiple detections. \citet{2017MNRAS.471..210G} present a 5.5 GHz sample of 94 sources (76 of which also present in our sample) in GOODS-N as part of the eMERGE survey, with a sensitivity of 3 $\mu$Jy/beam and $0.5\arcsec$ resolution. \citet{2017ApJ...839...35M} present 10 GHz high-resolution (FWHM $\sim 0.22\arcsec$) observations reaching a sensitivity of 5.72 $\mu$Jy/beam, detecting 32 sources in the field, 22 of which are in our catalog.

\subsection{UV-to-FIR data}
The search for counterparts of the radio selected sources is based on the CANDELS/SHARDS catalog presented in \citet[][B19 hereafter]{2019ApJS..243...22B}, complemented by the {\it super-deblended} catalog of \citet{2018ApJ...853..172L} for the FIR photometry (from 24$\mu$m to 1.1mm), and SCUBA-2 observations at 850$\mu$m \citep{2017ApJ...837..139C}.

The catalog from B19 is F160W-selected (1.6 $\mu$m) in the COSMOS and GOODS fields and built by assembling various ancillary multiwavelength observations, from the UV to the far-IR \citep{2003sptz.prop..196D, 2011ApJS..197...35G,koekemoer2011,2011PASJ...63S.379K, 2013ApJ...769...80A, 2015ApJS..218...33A, 2019ApJ...871..233H}. The WFC3 maps were obtained with a stratified ({\it ``wedding cake''}) observing strategy. The maps reach a detection limit at 5$\sigma$ of H = 27.8, 28.2 and 28.7 going from the outermost part of the maps to the  deep central regions, covering 50\%, 15\% and 35\% of the total GOODS-N area, respectively. The final B19 catalog in the GOODS-N comprises 35,445 sources.

All the products of B19, including catalogs and maps, are publicly available\footnote{\href{http://rainbowx.fis.ucm.es/Rainbow_slicer_public/}{\url{http://rainbowx.fis.ucm.es/Rainbow_slicer_public}}}. We use the F160W-based catalog for counterpart association and photometry, and the available maps to infer the photometry (or the upper-limit rms) of those radio-galaxies which are undetected. The B19 sample is constructed by running \textsc{SExtractor} \citep{bertin1996} in dual mode with the F160W-catalog as a prior, therefore in shallower bands (e.g. the CFHT $K$-band) we end up by having detections for some sources which are fainter than the nominal 5$\sigma$ limiting magnitude (see Sec.\,\ref{subsec:cpart} and Tab\,\ref{tab:data})

At wavelengths longer than 8.0 $\mu$m we do not use the B19 photometry, which is based on a simple cross-matching of Spitzer 24$\mu$m \citep{2005ApJ...630...82P} and Herschel \citep{2011A&A...532A..49B, 2011A&A...532A..90L, 2012MNRAS.424.1614O, 2013A&A...553A.132M} catalogs. Instead, our FIR photometry comes from the {\it super-deblended} catalog presented in \citet{2018ApJ...853..172L}, which uses prior positions from deep Spitzer 24$\mu$m and VLA 20cm data in order to fit the FIR/submm data. As such, \citet{2018ApJ...853..172L} are able to recover more sources, by pushing the FIR detections to lower thresholds and Herschel fluxes. Filters from from Spitzer IRS/PUI at 16$\mu$m to 1.1mm observations carried with AzTEC+MAMBO (whose catalogs are not included in B19) are listed in Tab.\,\ref{tab:data}. Out of the 2626 sources within the HST/WFC3 footprint  we only keep those with a reported total infrared ${\rm S/N} \geq 5$\footnote{The total infrared signal-to-noise ratio is defined as the quadrature sum of the S/N ratios measured in all the bands.}, following the advice given by \citet{2018ApJ...853..172L}. We also add 850 $\mu$m fluxes from SCUBA-2 observations \citep{2017ApJ...837..139C}: 130 radio sources fall within the footprint.

In Tab.\,\ref{tab:data} we report a summary of all the bands used throughout this work.

\begin{table*}
    \centering
    \caption{Summary of all the filters and catalogs considered in this work. Sensitivities are given at 5$\sigma$; \% of $F_{\rm meas}$ refers to the percentage of the 554 sources identified in the F160W band with a counterpart in the different filters.}
    \label{tab:data}
    \begin{tabular}{lccccccc}
    \hline
    Instrument & Wavelength & Sensitivity (radius [$\arcsec$])  & \% of $F_{\rm meas}$ & Reference  \\
    \hline
    KPNO\_U     & 3593 \AA      & 26.7 (1.26) & 93.9 & (1) \\
    ACS/F435W   & 4318 \AA      & 27.1 (0.10) & 86.5 & (1) \\
    ACS/F606W   & 5915 \AA      & 27.7 (0.10) & 94.4 & (1) \\
    ACS/F775W   & 7693 \AA      & 27.2 (0.11) & 95.7 & (1) \\
    ACS/F814W   & 0.81 $\mu$m   & 28.1 (0.11) & 96.2 & (1) \\
    ACS/F850LP  & 0.90 $\mu$m   & 26.9 (0.11) & 96.4 & (1) \\
    WFC3/F105W  & 1.01 $\mu$m   & 26.4 (0.18) & 83.2 & (1) \\
    WFC3/F125W  & 1.25 $\mu$m   & 27.5 (0.18) & 98.4 & (1) \\
    WFC3/F140W  & 1.39 $\mu$m   & 26.9 (0.18) & 69.5 & (1) \\
    WFC3/F160W  & 1.54 $\mu$m   & 27.3 (0.19) &100.0 & (1) \\
    MOIRCS Ks   & 2.13 $\mu$m   & 24.7 (0.60) & 67.9 & (1) \\
    CFHT K      & 2.16 $\mu$m   & 24.4 (0.60) & 99.6 & (1) \\
    IRAC/CH1    & 3.56 $\mu$m   & 24.5 (1.7) & 99.3 & (1) \\
    IRAC/CH2    & 4.51 $\mu$m   & 24.6 (1.7) & 99.5 & (1) \\
    IRAC/CH3    & 5.76 $\mu$m   & 22.8 (1.9) & 98.4 & (1) \\
    IRAC/CH4    & 8.0  $\mu$m   & 22.7 (2.0) & 98.6 & (1) \\
    IRS16       & 16.0 $\mu$m   & 38.5 $\mu$Jy & 78.3 & (2) \\
    MIPS24      & 24.0 $\mu$m   & 26 $\mu$Jy & 84.7 & (2) \\
    PACS100     & 100  $\mu$m   & 1.6 mJy    & 84.7 & (2) \\
    PACS160     & 160  $\mu$m   & 3.4 mJy    & 84.7 & (2) \\
    SPIRE250    & 250  $\mu$m   & 7.85 mJy    &  83.4 & (2) \\
    SPIRE350    & 350  $\mu$m   & 10.35 mJy    &  69.9 & (2) \\
    SPIRE500    & 500  $\mu$m   & 12.58 mJy    & 38.4 & (2) \\
    SCUBA850    & 850  $\mu$m   & 1.65 mJy    & 13.7 & (3) \\
    AzTEC+MAMBO & 1160 $\mu$m   & 3.3 mJy    & 29.6 & (2) \\
    VLA         & 10 GHz        & 5.7 $\mu$Jy beam$^{-1}$ & 4.0  & (4) \\
    VLA         & 5.5 GHz       & 3.0 $\mu$Jy beam$^{-1}$ & 13.5  & (5)  \\
    VLA         & 1.4 GHz       & 2.2 $\mu$Jy beam$^{-1}$ & 100.0 & (6) \\
    \hline
    \end{tabular}
    \tablecomments{
 (1)~\citet{2019ApJS..243...22B}: 5$\sigma$ depth computed in apertures with radii reported in Column 3; (2)~Values derived as 5$\sigma$, where $\sigma$ is the flux uncertainty from ~\citet{2018ApJ...853..172L}; (3)~Values derived as 5$\sigma_{s}$, where $\sigma_{s}$ is the statistical uncertainties from ~\citet{2017ApJ...837..139C} (i.e. not considering confusion); (4)~\citet{2017MNRAS.471..210G}; (5)~\citet{2017ApJ...839...35M}; (6)~\citet{2018ApJS..235...34O}
 }
\end{table*}

\subsection{Ancillary redshifts}
B19 provides redshifts for most of the sources presented in their sample. These redshifts fall into the usual categories: $z_{\rm spec}$ for sources with spectroscopic redshift, the great majority coming from \citet{2008ApJ...689..687B}, and $z_{\rm phot}$ for sources whose redshifts have been derived from their best-fit SED. The latter category is subsequently divided in three tiers: $z_{\rm phot}$ evaluated with broadband filters only, $z_{\rm phot}$ evaluated with broadband filters plus SHARDS photometry \citep{2013ApJ...762...46P}, $z_{\rm phot}$ evaluated with broadband filters and/or SHARDS photometry plus WFC3-grism  ($z_{\rm grism}$). These last ones have a reliability flag higher with respect to photometric redshifts, though lower than spectroscopic ones. In this work, we adopt the B19 $z_{\rm spec}$ and the ones belonging to the latter category, and re-evaluate the remaining $z_{\rm phot}$.

For two sources we adopt the spectroscopic redshifts of 1.7844 \citep[ID. 386,][]{2014ApJ...782...78D} and 1.9200 \citep[ID. 753,][]{2015A&A...577A..46D}, both measured from multiple sub-mm CO lines detection, which we consider more reliable with respect to the previously reported redshifts, based on the identification of a single emission line in the NIR. 

\section{The sample}\label{sec:Sample}
We focus our analysis on the 578 radio sources from the \citet{2018ApJS..235...34O} catalog inside the F160W footprint, (see Fig.\,\ref{fig:coverage}), as stated in Sec. 2.1.

\subsection{Counterpart association}\label{subsec:cpart}
\begin{figure}
    \centering
    \includegraphics[width=1.0\columnwidth]{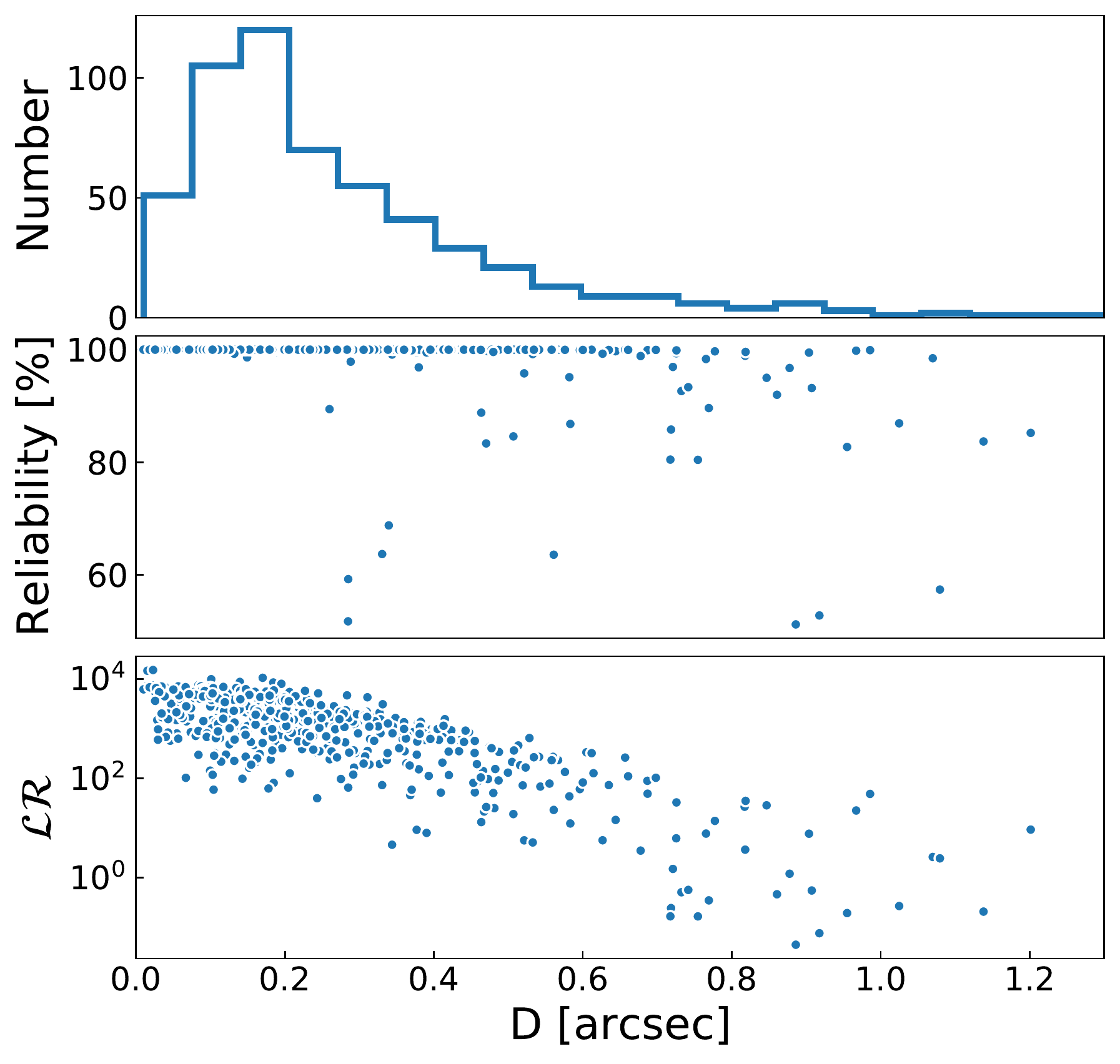}
    \caption{Distribution of the distance between the likelihood-ratio chosen counterparts from the B19 catalog and our radio sources (upper panel). Also reported the reliability in percentages (middle panel), and the likelihood ratio (lower panel).}
    \label{fig:LRacc_distances}
\end{figure}

We associate counterparts to the radio sources following the likelihood ratio (LR) technique presented in \citet{2003A&A...398..901C, 2018A&A...620A..11C}. The likelihood ratio is defined as the ratio between the probability that a given source is the true counterpart and the probability that the same source is an unrelated background object. The choice of one source among the others is expressed in terms of {\it reliability}, given as a percentage, as in Eq. 5 of \citet{1992MNRAS.259..413S}
\begin{equation}
    {\rm Rel}_j = \frac{{\rm LR}_i}{\Sigma_i {\rm LR}_i + (1 - Q)}
\end{equation}
where the $i$ sum is intended over the possible candidate counterparts for each radio source. The source with the highest LR will have the highest reliability, and as such is the associated counterpart.

Starting from our sample of 578 radio sources, within a 1.6$\arcsec$ radius (equal to the resolution of the radio map), we obtain a reliable association for 548 objects (see Fig.\,\ref{fig:LRacc_distances})

We also use the likelihood ratio technique to link the radio sources to the ones in the \citet{2018ApJ...853..172L} and \citet{2017ApJ...837..139C} catalogs: 560 sources have a FIR association within 2.0$\arcsec$, and 90 have a SCUBA-2 counterpart within 3.6$\arcsec$.

For all the 548 radio sources with an associated NIR counterpart we visually check the cutouts, in order to assess the quality of the association, and correct possible issues. We found that the vast majority of associations are good, but in a couple of cases we manually fixed them and chose the most likely source by eye. For example, in one case the radio emission falls exactly between two faint NIR sources, while in another case the likelihood ratio slightly privileged a larger and brighter source at approximately 1$\arcsec$ over a fainter one whose emission center is almost perfectly coincident with the radio one.

The visual check is similarly performed on the 30 sources without an associated F160W counterpart. In one case, the extended radio emission clearly fell over a strong NIR emission, which is however separated by more than 1.6$\arcsec$ from the center of the radio emission. Five objects are very extended low-$z$ sources with the $H$-band emission center farther than 1.6$\arcsec$ from the radio emission center. We added these sources to our sample of radio galaxies with an $H$-band association, which finally counts 554 objects.

As for the remaining 24 sources, in three cases the radio emission falls clearly on a source which is not present in the B19 catalog either because at the border of the F160W map or due to extreme contamination from a strong nearby source. For other four sources the radio emission is barely above the 5$\sigma$ level, either blended into multiple knots of emission or falling in between three NIR sources at $>3\arcsec$. These seven sources are conservatively removed from the sample. 
The residual outcome of the counterpart association and visual check is a group of 17 sources that we call $H$-dark. Three of these  actually show a marginal detection (below the 3$\sigma$ level).
We point out that the label of $H$-dark galaxies is not an absolute definition and it depends of course on the specific depth of the F160W map in the GOODS-N that we are considering in this work.

We search for counterparts to the $H$-dark galaxies in the IRAC bands by looking at the maps and we find 8 counterparts out of 17. Five sources are well isolated in the field, with minimal, if any, contamination from nearby sources, therefore we are able to measure their IRAC fluxes with an aperture-corrected photometry of 2$\arcsec$ radius performed with the {\sc photutils} package of {\sc astropy} \citep{larry_bradley_2019_3568287}. The IRAC emission of the other three galaxies is slightly contaminated by a nearby source, hence the photometry is evaluated with \textsc{Tractor} \citep{2016ascl.soft04008L, 2016AJ....151...36L, weaver2021}, a \textsc{Python} module for image modeling that uses the known priors on sources position to fit the observed fluxes in multiwavelength bands with a variety of different models, effectively deblending the emissions coming from different sources.

To recap, our final sample consists of 554 sources with an association in the B19 catalog, and 17 $H$-dark sources.

\subsection{SED fitting and photometric redshifts}

The next step is the characterization of our sources through SED fitting. This is done differently depending on the availability of a reliable spectroscopic redshift or on the necessity to also estimate a photometric redshift.

B19 provide redshifts for every source in their catalog, either coming from a number of different spectroscopic surveys ($z_{\rm spec}$) or a SED-fitting evaluated photometric redshift ($z_{\rm phot}$). Of the 554 radio sources with a NIR counterpart 392 ($\sim 70\%$) have a reliable spectroscopic redshift and the remaining 162 ($\sim 30\%$) have a photometric redshift.

The photometric redshift evaluation in B19 is done using six different codes ({\sc EAZY, HyperZ, SpeedyMC, Lephare, zphot, WikZ}) and taking the median of the multiple photo-$z$ estimates as the best-fit value. We note that these codes use galaxy emission models extending up to MIR photometry. As reported in Sec.\,\ref{sec:Data}, out of the three tiers of photometric redshifts we keep those estimated also with WFC3 grism data and re-evaluate the photometric redshifts for the remaining 90 sources. We use only the broadband photometry but also add the FIR photometry from the {\it super-deblended} catalog.

The photo-$z$ measurement procedure is performed using the {\sc magphys+photoz} code \citep{2019ApJ...882...61B, 2015ApJ...806..110D}. \textsc{magphys} models the emission in the entire UV-to-FIR range assuming that the energy output is balanced between the emission at UV-to-NIR wavelengths and the one absorbed by dust and then re-emitted in the FIR. It follows a Bayesian approach to measure the posterior distribution functions (PDF) of the parameters, fitting the observed photometry to a set of galaxy emission models coming from native libraries, composed of 50000 stellar population spectra with $\psi (t) \propto \exp^{- \gamma t}$ star formation histories with superimposed random bursts \citep{2003MNRAS.344.1000B}, associated via the energy balance criterion to 50000 dust emission SEDs with two components at different temperature \citep{2000ApJ...539..718C, 2008MNRAS.388.1595D}. In the \textsc{magphys+photoz} extension, the code displaces these models in a wide grid of redshifts, to infer the PDF of the galaxy photo-$z$.

In the cases of non detections we choose to adopt 3$\times$rms as upper limits, in order to allow the code to explore the SED models space more freely. We stress that the MAGPHYS does not force the SED model to stay strictly below the limit, instead it treats the upper limits as values with zero flux and error equal to the upper-limit value (see Sec 4.2.3 in the \textsc{magphys+photoz} documentation). We also verified that there is no significant difference in the  photo-z values obtained by adopting 1$\times$rms upper limits instead of 3$\times$rms.

\begin{figure*}
    \centering
    \includegraphics[width=1.0\columnwidth]{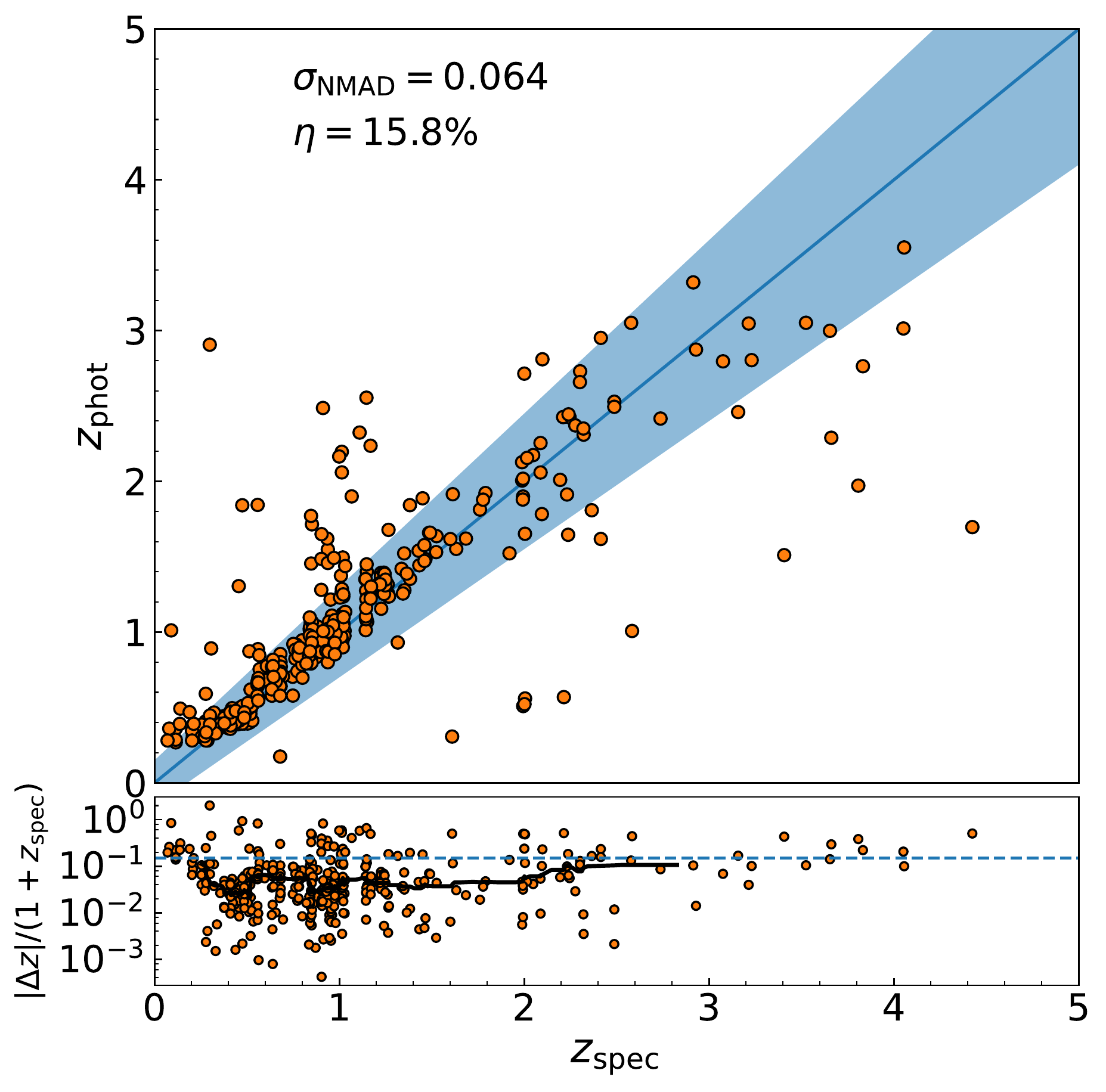}
    \includegraphics[width=1.0\columnwidth]{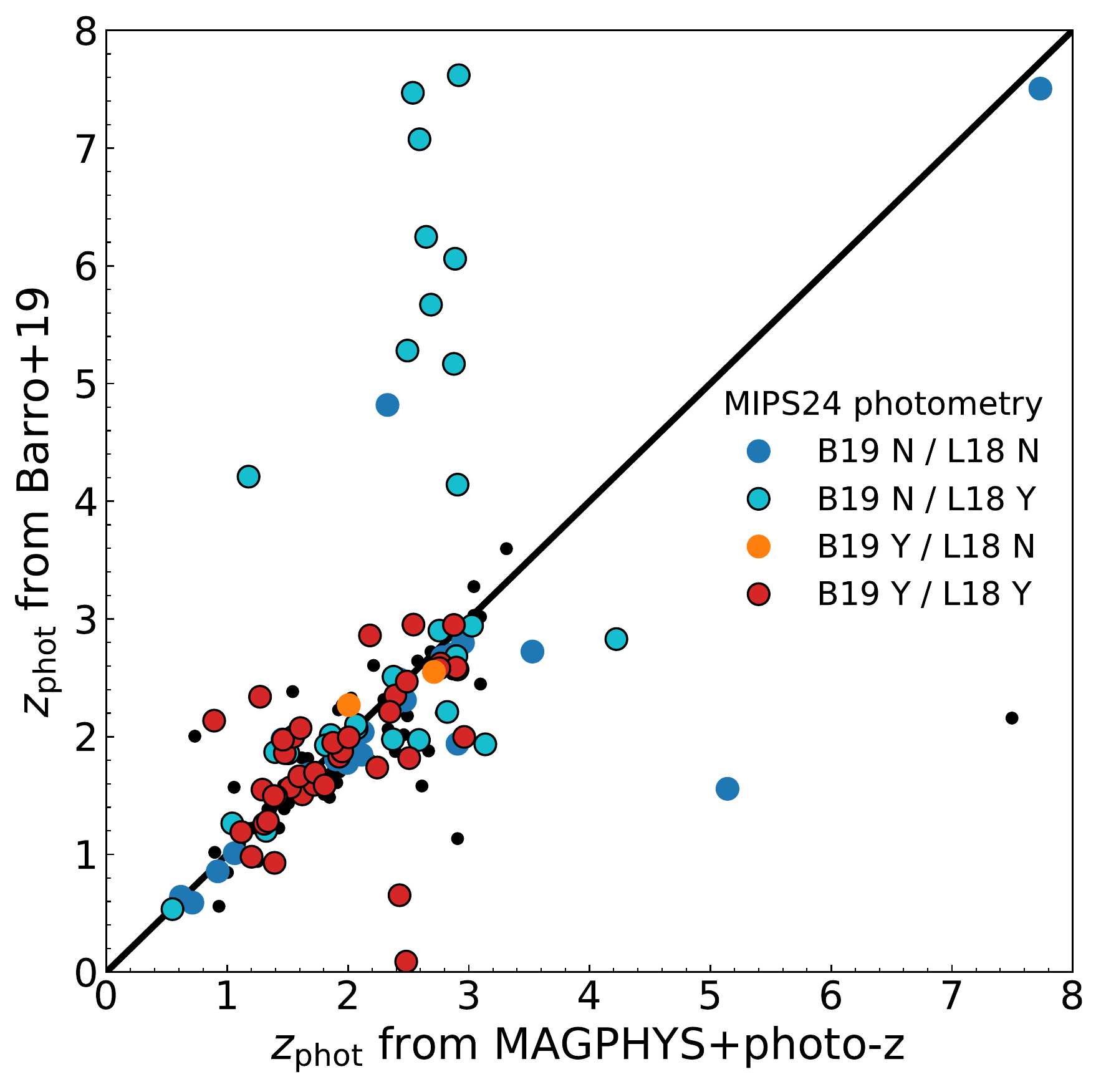}
    \caption{({\it Left panel}): Comparison between $z_{\rm spec}$ as reported in B19 and $z_{\rm phot}$ evaluated with \textsc{magphys+photoz}. $\sigma_{\rm NMAD}$ is the normalized median absolute deviation of the distribution, while $\eta$ is the fraction of outliers (defined as the sources with $\eta = \left| \Delta z \right| /(1 + z_{\rm spec}) > 0.15$), above the blue dotted line in the bottom panel and outside the blue shades around the 1:1 relation in the top panel. Black line is the running median. ({\it Right panel}): Comparison between the $z_{\rm phot}$ reported in B19 and the $z_{\rm phot}$ evaluated with \textsc{magphys+photoz} for the 90 sources for which we re-evaluated the redshift. The presence (red/orange) or absence (blue/cyan) of far-infrared photometry in the B19 catalog is highlighted with different colors, while the presence in our catalog (taken from \citet{2018ApJ...853..172L}, L18) is highlighted with a marker border. There is a good agreement between the two redshifts, with only 14/90 ($\sim 16\%$) of outliers, the vast majority belonging to sources without FIR in B19. We also show (small black points) the 72 sources labeled as $z_{\rm grism}$ (for which we adopt the B19 redshift) in order to illustrate the comparison with \textsc{magphys+photoz}.}
    \label{fig:z_comparison}
\end{figure*}

In order to take into account the possible presence of nuclear activity which could give a non negligible contribution to the SED and bias the radio luminosity, we also use \textsc{sed3fit} \citep{2013A&A...551A.100B}, which is a \textsc{magphys}-inspired SED fitting code, accounting for mid-IR emission possibly coming from the dusty torus of an AGN \citep{2012MNRAS.426..120F}. Since \textsc{sed3fit} does not implement photometric redshifts measurement, we fix it to the ones coming from \textsc{magphys+photoz}, for the sources for which we re-evaluate it.

A quality assessment of \textsc{magphys} photo-$z$ is done measuring photometric redshifts for the subsample of galaxies with available $z_{\rm spec}$ and looking at the outliers fraction of the spectroscopic vs. photometric distribution, shown in the left panel of Fig.\,\ref{fig:z_comparison}. 

In the right panel of Fig.\,\ref{fig:z_comparison} we compare the photometric redshifts of B19 to the ones obtained with \textsc{magphys+photoz} for the first two tiers of redshift reported in B19 (i.e. without the $z_{\rm grism}$, 90 objects). Not every source in B19 has MIPS photometry at 24$\mu$m, while we have a full far-IR photometry taken from the {\it super-deblended} catalog from \citet{2018ApJ...853..172L}. As such, we highlight the presence (or lack) of far-IR data using different colors and marker edge colors. There is a general agreement between the two redshifts, with a fraction of outliers around $16\%$ of the sample. All but two of the outliers are sources for which we have FIR photometry from the superdeblended catalog and B19 do not. The presence of photometric measurements in the mid-IR (in particular the IRS16 and MIPS24 channels sampling the PAH emissions) likely allows us to obtain a better constraint for photo-$z$.
We note that our photometric redshifts are in good agreement with the ones that we label as grism-based redshift.

We also apply the described SED-fitting procedure to the 8 $H$-dark galaxies with an associated IRAC emission. At wavelengths shorter than IRAC1, as done for the rest of the sample, we assume as upper limits 3$\times$rms, measured directly on the maps, with the exceptions of two sources for which $K_s$ photometry is available. 

The best-fit photometric redshifts of the $H$-dark galaxies lie in the $3.00 < z_{\rm phot} < 3.25$ range, with the exception of one source at $z=1.74$ and one at $z=4.23$, the latter showing quite a broad $z$-PDF (see Figure \ref{fig:SEDs} in the Appendix). 

As for the remaining 9 $H$-dark sources the photometry is either not sufficient or too heavily contaminated by nearby sources to construct a robust SED and estimate a photometric redshift.

In Fig.\,\ref{fig:logLvsz} we show the redshift distribution of our sample.

\section{The 1.4-GHz luminosity function}\label{sec:lumfunc}
In order to measure the evolution of cosmic star formation, and put constraints on the contributions of the different populations, we first need to estimate the statistical properties of our sample per unit of Universe comoving volume.

The radio luminosity function (LF, $\phi$) is defined as the number density of radio sources in a given volume of Universe per luminosity bin. When measured in different redshift bins, the LF gives information on the density and luminosity evolution of the sources from which it is built. The conjunction of known relations between the luminosities and physical properties (e.g. $L \rightarrow \mathcal{SFR}$) with its integration in a certain luminosity range returns insight on processes such as the star formation undergoing at a given cosmic epoch.

Our sample's redshifts go up to $z \sim 4.5$. In principle, there is a source (ID.424) at $z$-phot $\sim 7.7$ but we exclude it from the LF evaluation, since a single source is not representative of the physical condition of the Universe at those extremely high-$z$.

Starting from the assumption that the radio emission follows a simple power-law $S_{\nu} = \nu^{\alpha}$, and therefore the radio $K$ correction is $K(z) = (1+z)^{-(1+\alpha)}$, the final expression for the radio luminosity $L$ at rest-frame frequency of 1.4 GHz derived from the observed flux density $S_{\nu}$ can be written as
\begin{equation}
L_{\rm 1.4 GHz} = \frac{4 \pi D_L^2(z)}{(1+z)^{1+\alpha}} S_{\nu}
\end{equation}
with $D_L$ being the luminosity distance at source redshift $z$.

\begin{figure}
    \centering
    \includegraphics[width=1.0\columnwidth]{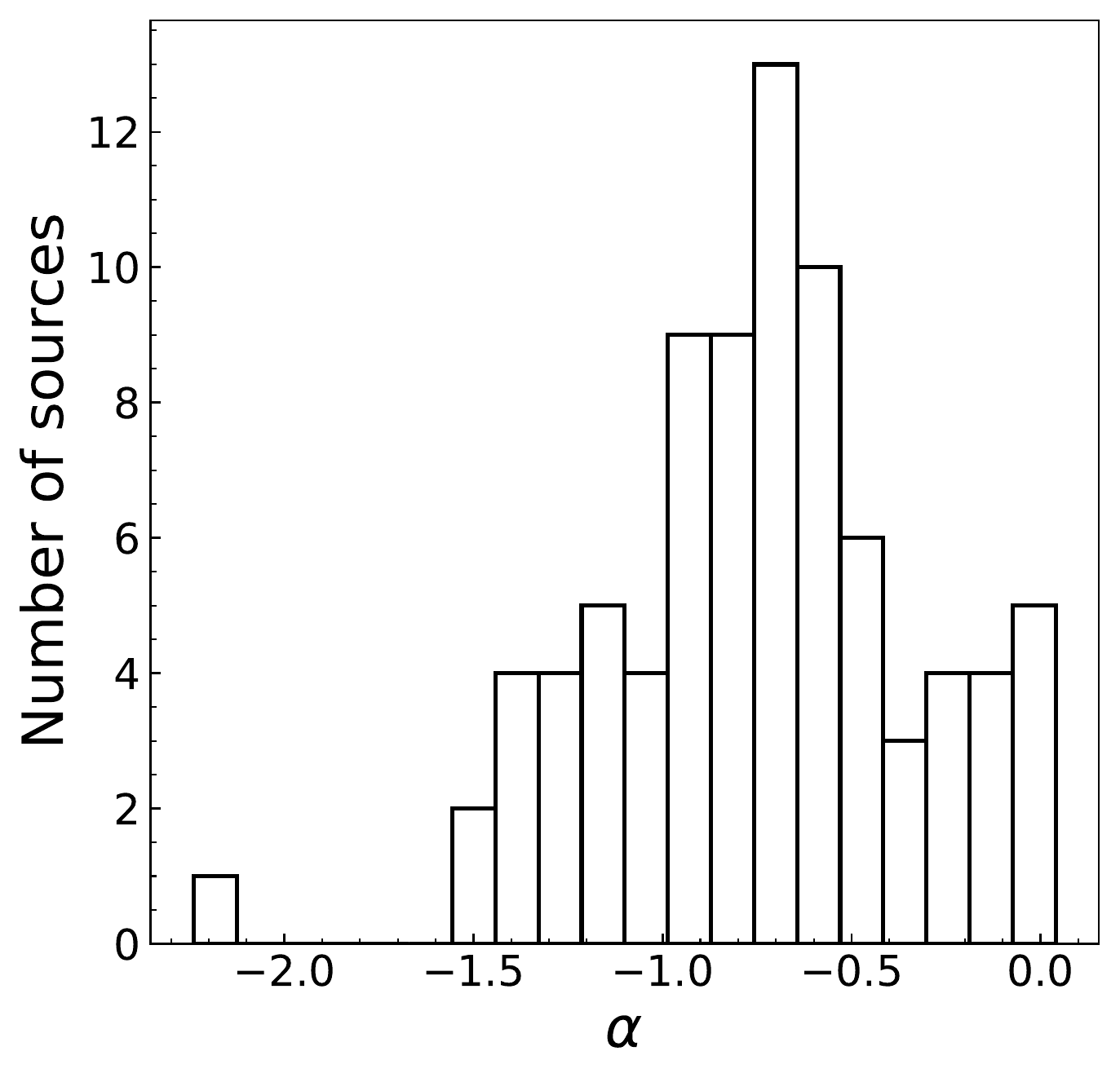}
    \caption{Best-fit radio spectral index $\alpha$ for the 76 galaxies with multiple detections in \citet{2017ApJ...839...35M, 2017MNRAS.471..210G, 2018ApJS..235...34O}. Median value is -0.73, $1\sigma$ dispersion 0.42. For the rest of the sample we assume a spectral slope of $-0.7$.}
    \label{fig:spectralslope}
\end{figure}

There are $79$ sources in our sample for which 5 GHz \citep{2017MNRAS.471..210G} and/or 10 GHz \citep{2017ApJ...839...35M} data exist (57 only have 5 GHz, 3 only 10 GHz, 19 with both 10 GHz and 5 GHz data) allowing a measurement of the power-law index. Their distribution is shown in Fig.\,\ref{fig:spectralslope}, with a median value of $-0.73$ and a dispersion of 0.41, consistent with the value of $-0.7$ usually adopted in the literature \citep[i.e.][]{2017A&A...602A...5N}. Taking this into account, we fix the power-law index to $-0.7$ for the sources for which we could not directly measure it.

\begin{figure*}
    \centering
    \includegraphics[width=0.9\textwidth]{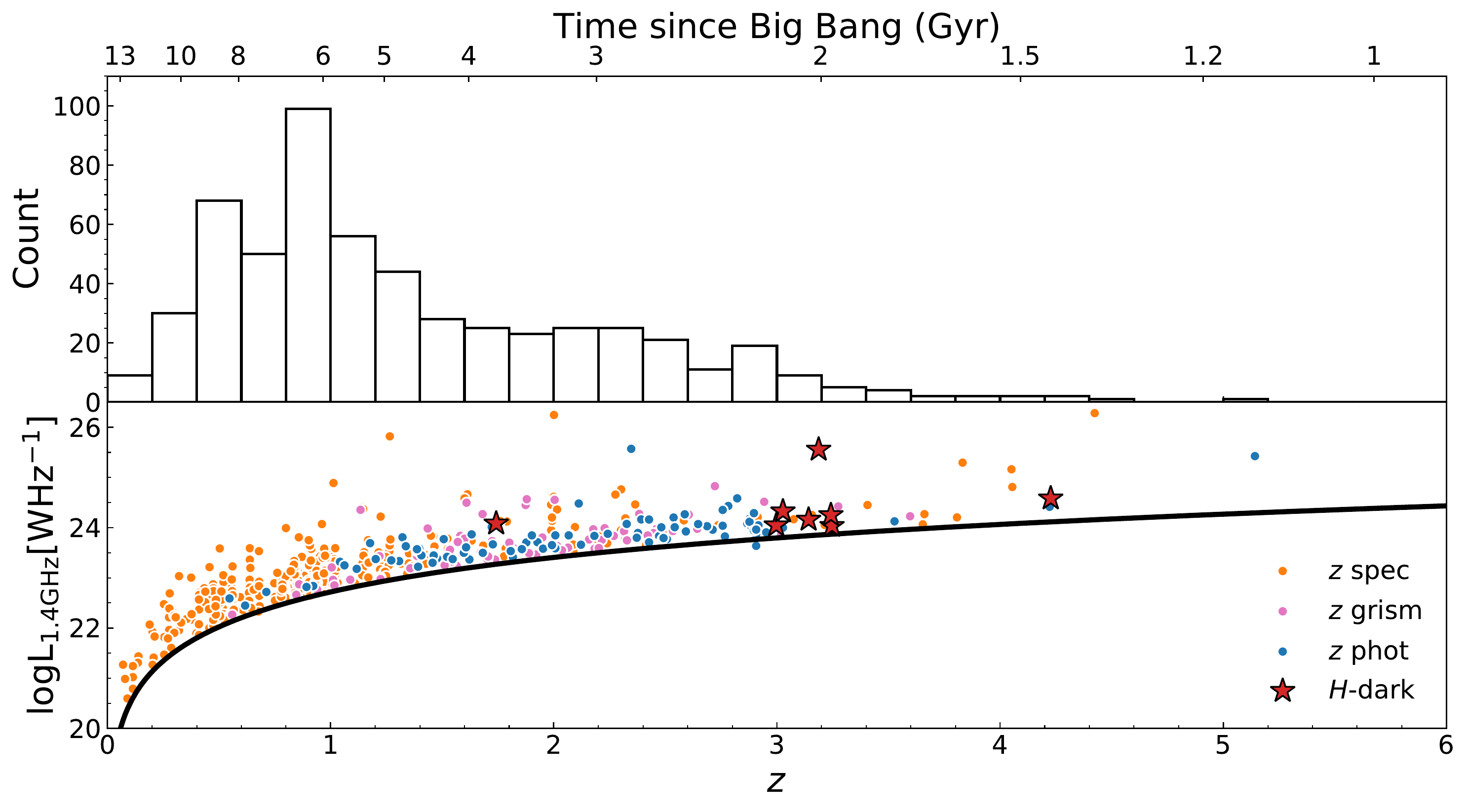}
    \caption{The redshift distribution of the sample {\it (top panel)} and rest-frame 1.4 GHz luminosity {\it (bottom panel)}. Sources are color-coded per redshift type, spectroscopic (orange), grism (pink), photometric (blue). $H$-dark galaxies are highlighted with red stars. Black line is the 5$\sigma$ detection limit obtained from the survey flux limit at the best-resolution image of 2.2 $\mu$Jy beam$^{-1}$ and a fixed spectral slope of $\alpha = -0.7$.}
    \label{fig:logLvsz}
\end{figure*}

The number counts and the rest-frame 1.4 GHz luminosity, color-coded per redshift type, are shown in Fig.\,\ref{fig:logLvsz}.
\begin{figure}
    \centering
    \includegraphics[width=1.0\columnwidth]{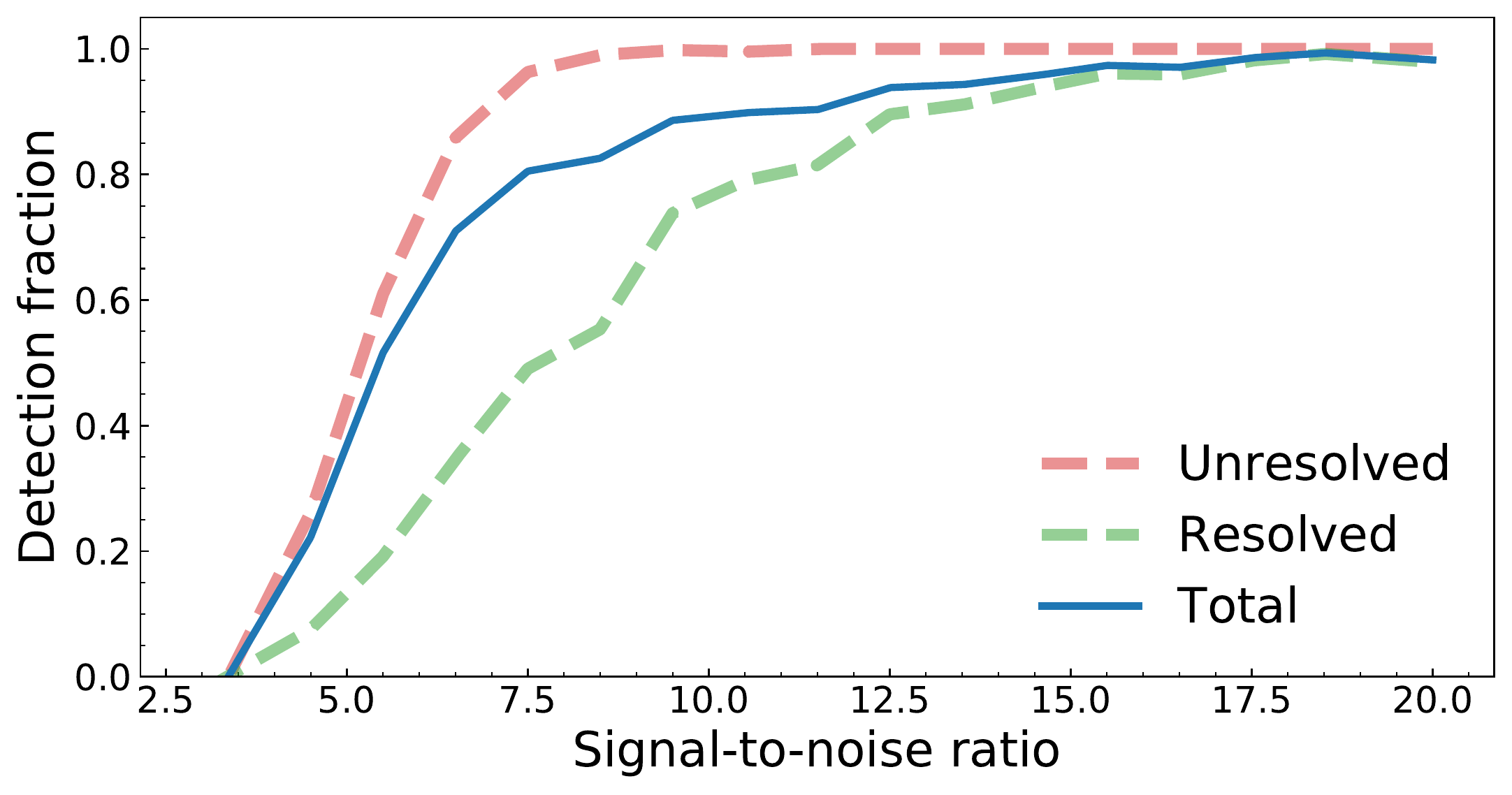}
    \includegraphics[width=1.0\columnwidth]{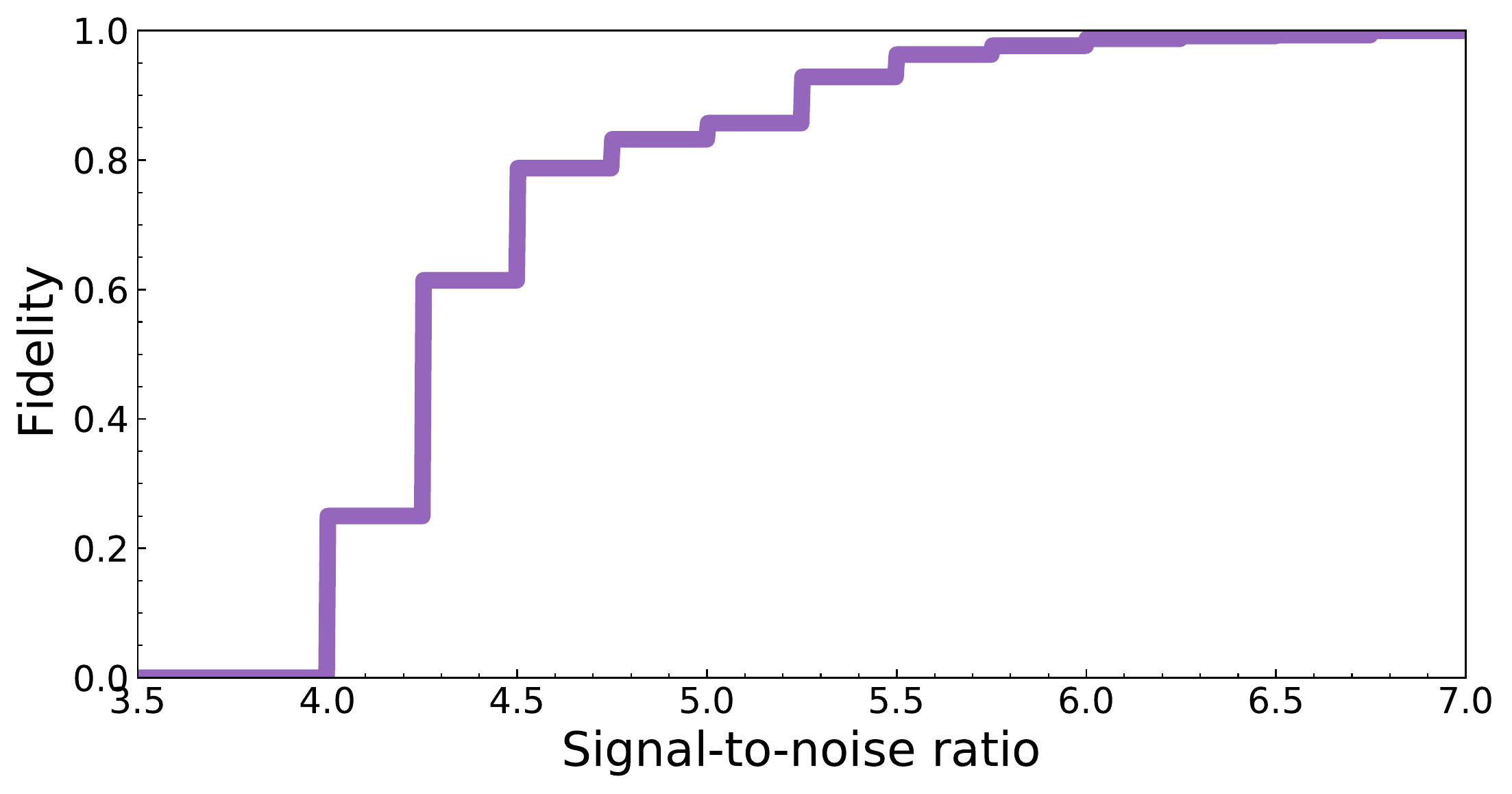}
    \caption{The sample detection fraction ({\it upper panel}) and fidelity ({\it lower panel}), both obtained from simulations with injected sources of known flux. {\it Upper panel}: blue line is the detection fraction for the full sample, while green and red dashed lines are for the resolved and unresolved injected sources, respectively. {\it Lower panel}: purple step function is the sample fidelity as a function of SNR.}
    \label{fig:detection_fraction}
\end{figure}

\subsection{Completeness correction and sample fidelity}\label{subsec:completeness}
In order to estimate the incompleteness of our sample, we inject artificial sources in simulated maps, and look at the fraction of recovered sources as a function of the flux \citep[such as e.g.][]{2017A&A...602A...1S, 2018A&A...620A..74R, 2020A&A...643A...2B}. 

We simulate multiple VLA observations of the same field, reaching the same noise level as the real observation, therefore assuming that the statistical properties of all the maps are the same. In each simulation, we inject $\sim 2000$ sources, separated by at least two times the beam size, whose fluxes are distributed as the observed source counts extrapolated (with a power-law) down to a flux density of 5$\mu$Jy, thus allowing us to investigate the detectability of sources below the 5$\sigma$ level. Source sizes come from a fit to the total-to-peak flux ratio $S_t/S_p$
\begin{equation}
    \frac{S_t}{S_p} = \frac{\sqrt{\theta^2_M + \theta^2_b} \sqrt{\theta^2_m + \theta^2_b}}{\theta_b^2}
\end{equation}
with $\theta_M$ and $\theta_m$ being the intrinsic source major and minor Full-Width at Half-Maximum (FWHM) angular sizes, and $\theta_b$ the beam FWHM. The simulated sources are built in order to mimic the observed relations between $S_t$, $S_t/S_p$ and $\theta_M$ (therefore $\theta_m$ as previously shown), in the various beam regimes from which the sample is extracted ($\theta = 1.6, 2.0, 3.0, 6.0, 12 \arcsec$, see Sec\,\ref{sec:Data}).

The injection is carried out directly on the visibilities, in order to take into account effects of the imaging process of bright sources, i.e. side lobes. Source extraction, and therefore the detection fraction, is done with the Python Blob Detector and Source Finder (PyBDSF), which uses the same Gaussian component fit and removal of the task SAD of AIPS used in O18 to build the catalog of detected sources. In accordance with O18 method of catalog construction, we repeat the source extraction in multiple realizations of the same map, degraded to beams of 2.0$\arcsec$, 3.0$\arcsec$, 6.0$\arcsec$, 12.0$\arcsec$. We checked that in the simulated sample the $S_t/S_p$ ratios are consistent with the ones coming from the observed sample, as a Kolmogorov-Smirnov test applied on both shows that they are not generated from the different distributions. This also holds true when dividing the samples in two regimes of high ($S_t > 40 \mu$Jy) and low flux ($S_t < 40 \mu$Jy).

The results are shown in the top panel of Fig.\,\ref{fig:detection_fraction}, with the detection fraction reported as a function of SNR. The blue solid line is the detection fraction measured for the full sample, while the green and red dashed lines respectively refer to the extended and the circular (i.e. unresolved) parts of the simulations. We assume that the sample is complete for SNR higher than 20 (fluxes higher than $\sim 40\mu$Jy), and we choose to limit our analysis to a completeness over 50$\%$, in order to avoid over-compensating for sources in completeness regimes where we would risk to correct for more sources than actually observed. This translates to an integrated SNR $> 5.20$ for unresolved sources and $> 7.68$ for resolved ones.

We also perform a fidelity analysis on the sample. Fidelity is a function of SNR and it represents how much we trust that a source with a certain SNR is actually a real detection and not a noise artifact. Following the approach of \citet{2016ApJ...833...69D}, we define the fidelity of a source at a given SNR as
\begin{equation}
    f ({\rm SNR}) = 1 - \frac{N_{\rm neg}}{N_{\rm pos}}.
\end{equation}
$N_{\rm neg}$ is the number of negative detections, that is the number of detections in the inverse of the simulated maps, and as such a proxy for all the blobs of pure noise mistakenly assumed as a good detection. In this way, we are able to estimate the percentage of false detections at any given value of SNR. We limit our analysis in the regimes where the fidelity is higher than 90\%, with SNR $> 5.20$.

\begin{figure}
    \centering
    \includegraphics[width=1.0\columnwidth]{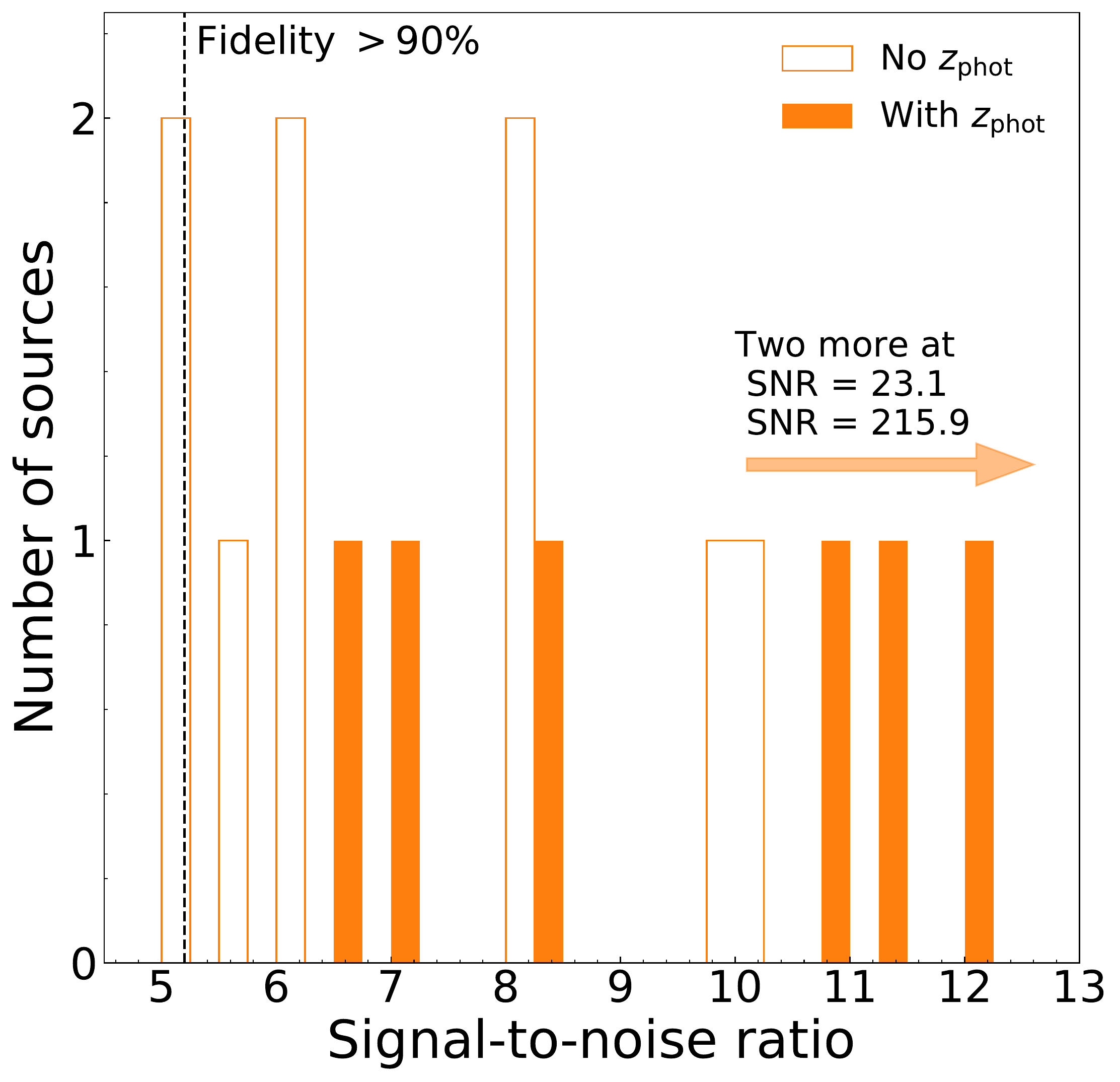}
    \caption{Signal-to-noise ratios at 1.4 GHz for the sample of 17 H-dark sources. Sources with SNR $> 5.2$ have a fidelity greater than 90\% (see Sec\,\ref{subsec:completeness}). Filled bars are for the sources with measured $z_{\rm phot}$.}
    \label{fig:Hdark_fidelity}
\end{figure}

In Fig\,\ref{fig:Hdark_fidelity} we show the SNRs for the 17 H-dark sources: all but two have 1.4 GHz SNR where the fidelity of the sample is over 90\%, and 14 of them belong to a SNR regime where fidelity is 100\% (SNR $> 6$). As such, we trust our sample of H-dark sources to be real with a high degree of confidence.

The SNR cuts reduce the pool of available sources to 479, i.e. 323 with $z_{\rm spec}$, 65 with $z_{\rm grism}$, 76 with $z_{\rm phot}$, and 15 $H$-dark sources, including the 8 with a measure of $z_{\rm phot}$. The luminosity function analysis in the following Section is derived considering this final selection.

\subsection{Estimating the AGN contamination}\label{subsec:AGNs}
Samples of radio-selected sources do not include only purely star-forming galaxies, as a fraction of their radio luminosities could be associated to nuclear activity. Disentangling the possible AGN contribution to the radio emission is necessary in order to properly convert luminosities into star formation rates.

This is usually achieved by conservatively removing those sources showing an excess in their 1.4 GHz emission due to nuclear activity \citep[{\it radio excess sources}, e.g.][]{2017A&A...602A...5N}. While this approach is successful in minimizing the AGN contamination, it completely removes from the sample SF-emission that could in principle be recovered, once disentangled from the AGN emission. In this work we tried to recover part of those sources by estimating the fraction of radio emission due to AGN ($f_{\rm AGN}$) and correcting the fluxes accordingly.

Measuring how an AGN can contribute to the radio emission goes through  the well-known infrared-radio correlation \citep[IRRC,][]{1971A&A....15..110V, 1985A&A...147L...6D, 1985ApJ...298L...7H}. The IRRC is a tight correlation with a dispersion $\sigma$ around 0.16-0.22 dex \citep{2017A&A...602A...3D, 2021arXiv210304803M}, linking the radio luminosity to the IR luminosity via the ratio
\begin{equation}
    q_{\rm TIR} = \log\frac{{\rm L}_{\rm IR} [{\rm W}] }{3.75\times 10^{12} [{\rm Hz}]} - \log \frac{{\rm L}_{1.4 {\rm GHz}}}{[{\rm W Hz^{-1}}]},
\end{equation}
where $3.75\times 10^{12} [{\rm Hz}]$ is the central frequency in the wavelength rest-frame 42-122 $\mu$m domain. In measuring $q_{\rm TIR}$ (and all the other related quantities) we use the AGN-removed IR luminosity provided by {\sc sed3fit}. With {\sc sed3fit}, we found a median mid-IR AGN fraction\footnote{Such fraction should not be confused with the $f_{\rm AGN}$ cited earlier, that is the fraction of AGN emission at 1.4 GHz.} of 16\% for the subsample of 81 galaxies with more than 5\% mid-IR AGN fraction, 12 of which over 50\%. 

Radio emission is the combination of free-free emission from \ion{H}{2} regions along with a major contribution of non-thermal emission coming from shock waves acceleration of relativistic electron produced in supernova explosions of massive stars, whereas the IR comes from the photon re-emission of dust heated by similarly massive OB stars, with the electrons experiencing different cooling processes (i.e. bremsstrahlung, inverse Compton scattering) in the host galaxies. The correlation holds in different environments \citep[merging or isolated galaxies,][]{1993AJ....105.1730C, 2002AJ....124..675C, 2013ApJ...777...58M} and over at least three orders of magnitude of luminosities \citep[e.g.][]{1985ApJ...298L...7H, 1992ARA&A..30..575C, 2001ApJ...554..803Y}.

Here, we use the IRRC, more specifically the offset from the IRRC, as a way to weight the AGN contribution in the radio \citep{2005ApJ...634..169D, 2013A&A...549A..59D, 2015MNRAS.453.1079B}. \citet{2017A&A...602A...3D} studied a sample of 7700 COSMOS radio sources, deriving an analytical expression for a (weakly redshift-dependent) $3\sigma$ threshold for sources dominated by AGN emission
\begin{equation}
r = \log\frac{L_{1.4 {\rm GHz}}[{\rm W Hz}^{-1}]}{{\rm SFR}_{\rm IR, SF}[M_{\odot} {\rm yr}^{-1}]} > 22 \times (1+z)^{0.0013}
\end{equation}
Sources above this ratio show a radio emission in excess with respect to the one compatible with pure star-formation processes. This criterion does not exclude all the possible AGN sources in a sample, but significantly reduces the amount of the ones whose emission is dominated by nuclear processes. Based on this criterion, there are 109 radio excess sources in our full sample ($\sim 20\%$), three of which are $H$-dark objects.
\begin{figure}
    \centering
    \includegraphics[width=1.0\columnwidth]{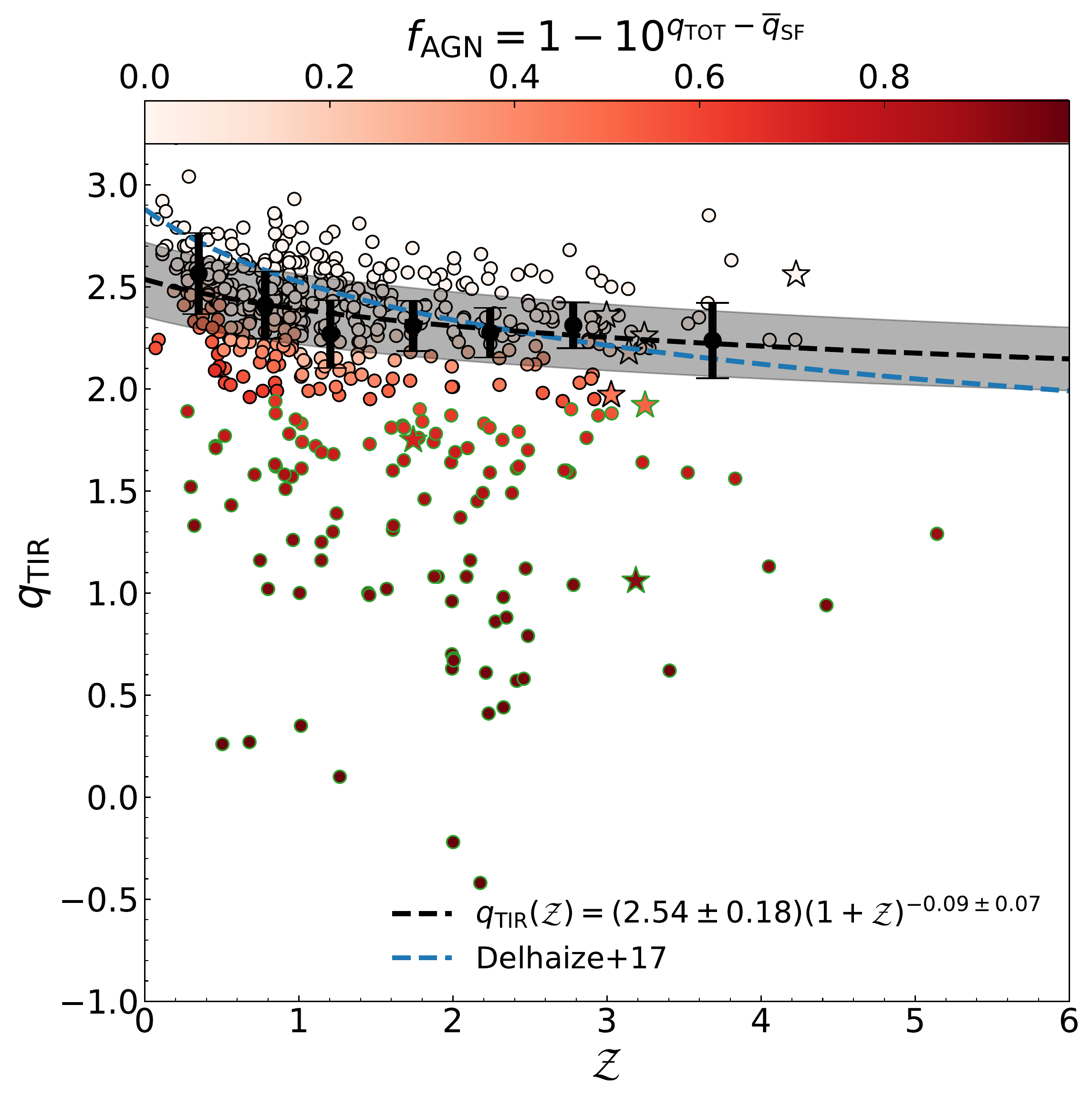}
    \caption{The IRRC for our sample as a function of redshift, color coded for $f_{\rm AGN}$. Black circles are measured in bins of $z$, evaluated as reported in the text, and the best-fit relation (black-dotted line) is reported in the legend, while the blue-dotted line is the \citet{2017A&A...602A...4D} relation. Galaxies plotted with a green outer circle are classified as radio excess sources following the criterion of \citet{2017A&A...602A...3D}. Star markers identify the $H$-dark.}
    \label{fig:qTIRvsz}
\end{figure}

More recently, \citet{2020arXiv201005510D} calibrated the IRRC considering both time and host stellar mass evolution, finding a strong correlation with $M_{\star}$ and a much weaker one with $z$ as in \citet{2017A&A...602A...3D}, suggesting that the often recovered $q_{\rm TIR}$ evolution with time is more a consequence of the detection of galaxies at lower $M_{\star}$ in the lower redshift bins than a real evolutionary feature.

\citet{2018A&A...620A.192C} built their radio AGN LF by measuring the fraction of AGN emission for each object from the IRRC, and correcting the observed radio emission accordingly with the measured $f_{\rm AGN}$. We apply the same method to disentangle the AGN emission from the one expected from the IRRC due to star-forming processes. We divide our sample into 7 redshift bins, locate the peak of $q_{\rm TIR}$ for each bin distribution ($\overline{q}_{\rm SF}$), mirror the right side of the distribution and fit the result with a Gaussian to obtain the bin error. We measure the AGN fraction as the difference between the galaxy own $q_{\rm TIR}$ and the distribution peak in the appropriate redshift bin $f_{\rm AGN} = 1 - 10^{q_{\rm TIR} - \overline{q}_{\rm SF}}$. We then fit the $\overline{q}_{\rm SF}$ values to infer the evolution of $q_{\rm TIR}$ with $(1+z)$, obtaining a weak trend with redshift almost compatible with no evolution at all ($-0.09 \pm 0.07$).

By multiplying the 1.4 GHz luminosities per the measured $f_{\rm AGN}$, we are making the assumption that the radio emission due to star-formation processes should in fact lie exactly on the IRRC. In our sample we identified 109 radio-excess galaxies. We completely remove from subsequent analysis 19 AGN-dominated sources ($f_{\rm AGN} > 95\%$). We also discard 44 galaxies that fall below the $5\sigma$ detection threshold after the $f_{\rm AGN}$ correction, since they would have not have been detected without the flux boosting due to the AGN presence. In the end, we retain 46 radio-excess sources, corrected for $f_{\rm AGN}$. The results are shown in Fig.\,\ref{fig:qTIRvsz}.

Finally, we point out that the results obtained removing the radio excess sources from the sample are consistent with the ones obtained from the $f_{\rm AGN}$-corrected sample. This is expected, since the total fraction of recovered AGNs that do not fall below the maps sensitivity limits is $\sim 9\%$ of the sample, which is not enough to translate into a systematic offset in the results. As mentioned at the beginning of this Section, in this work we report the results obtained from the $f_{\rm AGN}$-corrected sample.

\subsection{$1/V_{\rm max}$ method and fit}

\begin{figure*}
    \centering
    \includegraphics[width=1.0\textwidth]{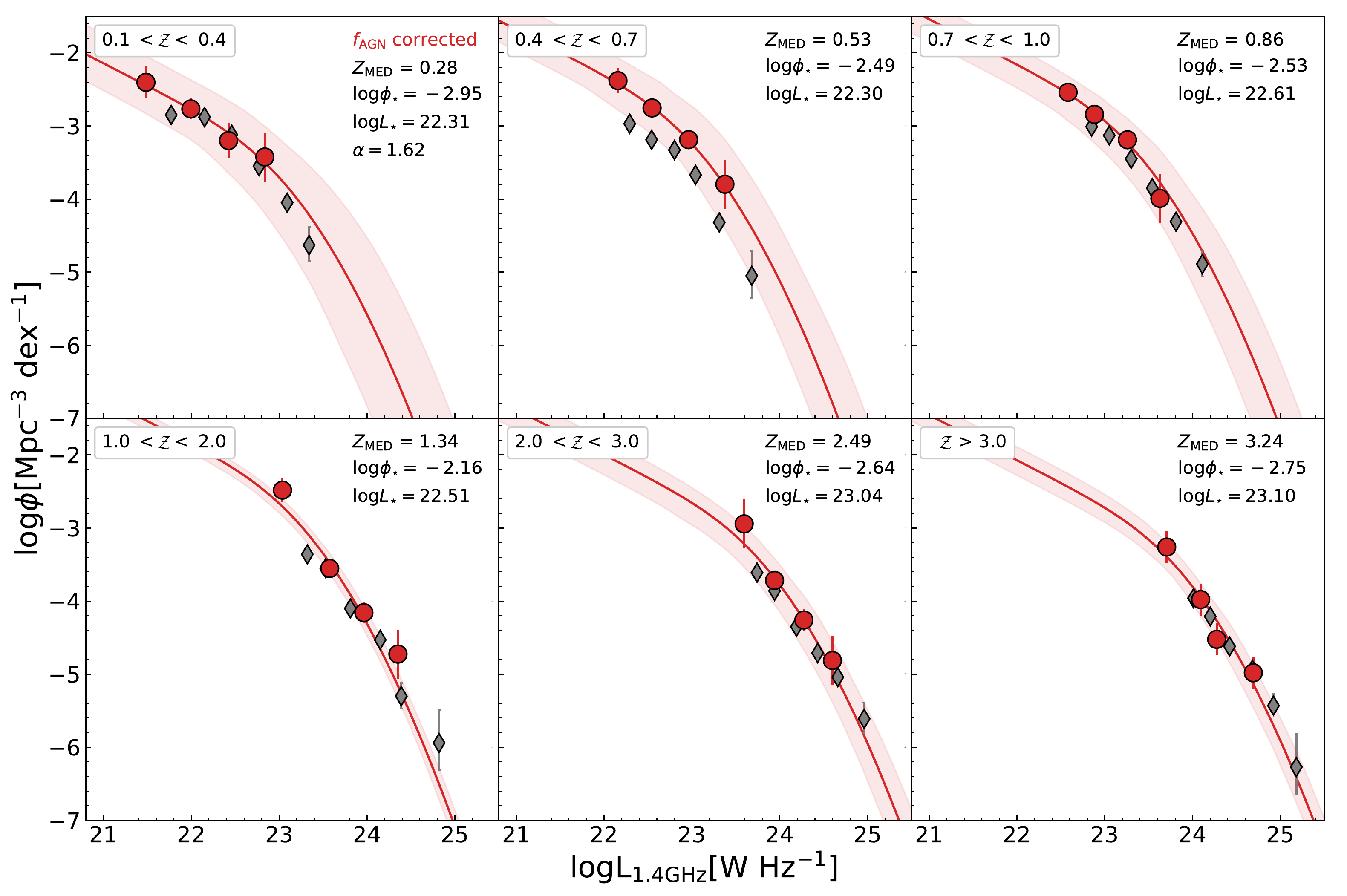}
    \caption{The 1.4 GHz luminosity function. Red circles are the measured $\phi$ luminosity function, while the line and shaded area are the best-fit and 1$\sigma$ curves, respectively. Grey diamonds are from \citet{2017A&A...602A...5N}.}
    \label{fig:LFs}
\end{figure*}

The 1.4 GHz radio LF is evaluated with the $1/V_{\rm max}$ method \citep{1968ApJ...151..393S}, a non-parametric approach that computes $\phi(L,z)$ measuring the source density from their maximum observable comoving volume, thus being completely independent from any analytical pre-assumption. The sum is performed on the sources falling within a certain redshift and luminosity bin, divided by the logarithmic bin width $\Delta\log L$:
\begin{equation}
\phi(L,z) = \frac{1}{\Delta\log L} \sum_i \frac{1}{V_{\rm max}}
\end{equation}

$V_{\rm max}$ is the maximum observable comoving volume of Universe for the $i$-th source, and is measured as the volume slice between the $z_{\rm min}$ of the redshift bin, and the minimum volume between the bin $z_{\rm max}$ and the redshift at which the source becomes undetectable given the survey depth. 

We consider the 472 sources with a measured redshift and a fidelity higher that 90 \% (including the 8 $H$-dark galaxies with photo-$z$, see Sec. \ref{subsec:completeness}).

Due to the specifics of the O18 sample construction, the survey depth varies in the five map realizations with different beams from which the sources are extracted, due to the different map noises. As such, we assign the proper detection limit to each source of the sample depending on the resolution map where the source is detected. Sources falling below this limit are excluded from the LF evaluation, i.e. 44 radio excess sources with $f_{\rm AGN}$-corrected luminosities.

We compute $V_{\rm max}$ by summing spherical shells in steps of $\Delta z = 0.01$ in the interval between $z_{\rm min}$ and $z_{\rm max}$ while correcting for the different biases affecting the sample:
\begin{equation}
V_{\rm max} = \sum_{z=z_{\rm min}}^{z_{\rm max}} \left( V(z+\Delta z) - V(z) \right) \times C_A C_I \left[ S_{1.4 \rm GHz} (z) \right]
\end{equation}
where $C_A$ corrects for the survey observed area of 171 arcmin$^2$
\begin{equation}
    C_A = \frac{A_{\rm survey}}{41253\,\,{\rm deg}^2}
\end{equation}
and $C_I \left[ S_{1.4 \rm GHz} (z) \right]$ is a statistical correction factor varying with the source flux as it moves through the redshifts, parametrising the survey incompleteness (see Sec.\,\ref{subsec:completeness}).

As in \citet{2017A&A...602A...5N}, sources with $\log L$ under the best-resolution detection limit at the bin $z_{\rm max}$ (black curve in Fig.\,\ref{fig:logLvsz}) are binned in a single luminosity bin, reducing the statistical impact of incompleteness in the faint end part of the LF. The subsequent bins are equally separated in $\log L$, with the exception of the last redshift bin, where due to the small number of sources we use bins containing almost the same number of sources.

In order to take into account the uncertainties on the $\sim 150$ sources with photometric or grism redshift, and the more general uncertainty on the radio luminosities, we measure the luminosity function for 1000 different realizations of $z$ and $\log L$, extracted from their probability density distributions. The final distributions are Gaussian, and as such the 50th percentiles of each distribution are taken for the luminosity bin centers and $\phi (L, z)$ points for each $\log L$ and $z$ bin.

The uncertainties on the luminosity function points are measured combining in quadrature the $1\sigma$ standard deviation of the aforementioned distributions and the error obtained weighting each galaxy contribution to the $V_{\rm max}$, following \citet{1985ApJ...299..109M}
\begin{equation}
    \sigma (L, z) = \frac{1}{\Delta \log L} \sqrt{\sum_i \frac{1}{V^2_{\rm max}}}
\end{equation}
In this way we account for the photometric uncertainties, though the fact that just a quarter of the sample has photo-$z$ and the large chosen redshift bins reduce the impact of these errors for most of the LF points, accounting for $\sim 10\%$ of the total error budget. Whenever there are less than 5 sources per luminosity bin, the uncertainty on the data point is calculated from the confidence intervals reported in \citet{1986ApJ...303..336G} for small number Poisson's statistics. In those cases, $\sigma (L, z) = \phi \times \sigma_N/N$.

The final LFs and their uncertainties are listed in Tab.\,\ref{tab:LF_values}, and reported as red dots in Fig.\,\ref{fig:LFs}.
\begin{table}
    \centering
    \caption{Luminosity function from the $1/V_{\rm max}$ method.}\label{tab:LF_values}
    \begin{tabular}{CCCCC}
    \hline
    {\rm Redshift} & \log L_{\rm 1.4 GHz} & \log \phi \\
    & \left[ {\rm W} {\rm Hz}^{-1}\right] & \left[ {\rm Mpc}^{-3} {\rm dex}^{-1}\right] \\
    \hline
    0.1 < z < 0.4 & 21.48 \pm 0.02 & -2.40^{+0.26}_{-0.70} \\
                  & 21.99 \pm 0.01 & -2.77^{+0.14}_{-0.14} \\
                  & 22.42 \pm 0.01 & -3.20^{+0.25}_{-0.73} \\
                  & 22.84 \pm 0.01 & -3.42^{+0.20}_{-0.87} \\
    0.4 < z < 0.7 & 22.16 \pm 0.03 & -2.38^{+0.17}_{-0.17} \\
                  & 22.55 \pm 0.00 & -2.75^{+0.07}_{-0.07} \\
                  & 22.96 \pm 0.01 & -3.19^{+0.12}_{-0.12} \\
                  & 23.38 \pm 0.02 & -3.80^{+0.20}_{-0.87} \\
    0.7 < z < 1.0 & 22.58 \pm 0.03 & -2.54^{+0.15}_{-0.12} \\
                  & 22.88 \pm 0.00 & -2.84^{+0.08}_{-0.07} \\
                  & 23.26 \pm 0.01 & -3.19^{+0.09}_{-0.09} \\
                  & 23.63 \pm 0.01 & -3.99^{+0.20}_{-0.87} \\
    1.0 < z < 2.0 & 23.04 \pm 0.06 & -2.48^{+0.16}_{-0.15} \\
                  & 23.57 \pm 0.01 & -3.55^{+0.07}_{-0.08} \\
                  & 23.96 \pm 0.03 & -4.16^{+0.14}_{-0.13} \\
                  & 24.35 \pm 0.06 & -4.72^{+0.20}_{-0.87} \\
    2.0 < z < 3.0 & 23.59 \pm 0.03 & -2.94^{+0.28}_{-0.31} \\
                  & 23.94 \pm 0.00 & -3.72^{+0.09}_{-0.11} \\
                  & 24.27 \pm 0.01 & -4.26^{+0.14}_{-0.16} \\
                  & 24.60 \pm 0.01 & -4.81^{+0.20}_{-0.87} \\
    3.0 < z < 5.2 & 23.70 \pm 0.05 & -3.26^{+0.26}_{-0.70} \\
                  & 24.09 \pm 0.03 & -3.98^{+0.26}_{-0.70} \\
                  & 24.27 \pm 0.03 & -4.52^{+0.26}_{-0.70} \\
                  & 24.69 \pm 0.01 & -4.98^{+0.26}_{-0.70} \\
    \hline
    \end{tabular}
\end{table}
\begin{table}
    \centering
    \caption{\textsc{ultranest} best-fit parameters to the measured luminosity function with the modified Schechter function in Eq.\,\ref{eq:Schechter}. $\alpha$ is measured in the first redshift bin, and fixed otherwise. $\sigma$ is fixed to 0.63.}\label{tab:LF_bestfit}
    \begin{tabular}{CCCC}
    \hline
    {\rm Redshift} & \alpha & \log L_{\star} & \log \phi_{\star} \\
    & & \left[ {\rm W} {\rm Hz}^{-1}\right] & \left[ {\rm Mpc}^{-3} {\rm dex}^{-1}\right] \\
    \hline
    0.1 < z < 0.4 & 1.62^{+0.56}_{-0.30} & 22.31^{+1.19}_{-1.05} & -2.95^{+0.71}_{-0.74} \\
    0.4 < z < 0.7 &                      & 22.30^{+0.98}_{-0.33} & -2.49^{+0.33}_{-0.70} \\
    0.7 < z < 1.0 &                      & 22.61^{+0.59}_{-0.28} & -2.53^{+0.29}_{-0.49} \\
    1.0 < z < 2.0 &                      & 22.51^{+0.18}_{-0.12} & -2.16^{+0.21}_{-0.28} \\
    2.0 < z < 3.0 &                      & 23.04^{+0.33}_{-0.28} & -2.64^{+0.47}_{-0.46} \\
    3.0 < z < 5.2 &                      & 23.10^{+0.30}_{-0.25} & -2.75^{+0.44}_{-0.50} \\
    \hline
    \end{tabular}
\end{table}

We fit the measured radio luminosity functions $\phi (L, z)$ in each redshift bin with a modified Schechter function \citep{1990MNRAS.242..318S}
\begin{equation}\label{eq:Schechter}
\phi_0 (L) = \phi_{\star} \left( \frac{L}{L_{\star}} \right)^{1-\alpha} \exp \left[ - \frac{1}{2\sigma^2} \log^2 \left( 1 + \frac{L}{L_{\star}} \right) \right]
\end{equation}
that behaves as a power-law function for luminosities below the turnover $L_{\star}$ and as a lognormal distribution for $L > L_{\star}$. This is a function of four parameters, $\alpha$ being the faint-end power-law index and $\sigma$ is determined by the bright-end, $L_{\star}$ the turnover-luminosity knee where the function behavior changes and ${\phi}_{\star}$ is the normalization. A modified Schechter is a typical assumption for the radio and IR luminosity function \citep[i.e.][]{1990MNRAS.242..318S, 2017A&A...602A...5N, 2020A&A...643A...8G}, accounting for more objects in the bright end part of the function than a normal Schechter function.

We assume that the shape of the LF does not change with cosmic time, and as such we fix the shape parameters $\alpha$ and $\sigma$, and measure $L_{\star}$ and ${\phi}_{\star}$. The bright end parameter $\sigma$ is fixed at 0.63, as in \citet{2017A&A...602A...5N}, where the authors infer the shape of the local LF from a sample of three radio surveys in the local Universe \citep{2002AJ....124..675C, 2005MNRAS.362....9B, 2007MNRAS.375..931M}. Given the depth of the radio observations upon which our sample is based, we are able to probe the faint end part of the LF only in the first redshift bin. As such, we evaluate the faint end $\alpha$ in the lowest-$z$ bin, and keep that measured value fixed for the other redshift bins.

Parameter space exploration, looking for the best-fit parameters, is performed with \textsc{ultranest}\footnote{\url{https://johannesbuchner.github.io/UltraNest/}}\citep{2021JOSS....6.3001B}, a Python module implementing importance nested sampling Monte Carlo algorithm MLFriends \citep{2014A&A...564A.125B, 2019PASP..131j8005B}. We assume flat priors for all the free parameters in the ranges $\log L{\star} : [19, 25]$, $\log \phi_{\star} : [-5, -1]$, $\alpha : [0, 3]$.

The measured luminosity functions and best-fit analytical forms are shown in Fig.\,\ref{fig:LFs}, where we also report the \citet{2017A&A...602A...5N} results in the closer redshift bin for comparison. LF values and best-fit parameters are reported in Tabs.\,\ref{tab:LF_values}-\ref{tab:LF_bestfit}.

For all the redshift bins, our LF reaches fainter luminosities by $\sim 0.2$ dex than the ones in \citet{2017A&A...602A...5N}, while they sample the bright end part better. The \citet{2017A&A...602A...5N} sample is derived from a wider area of 2 deg$^2$ in COSMOS, containing a factor $\sim 10$ more sources than our sample, which translates into LF points with smaller error bars, and access to brighter sources. The two LFs are in good agreement within the errors, with the notable exception of the bin $0.4 < z < 0.7$, where we obtain higher values at all luminosities, a difference that could be due to cosmic variance, i.e. an excess of galaxies in the redshift range corresponding to the bin.

\section{The cosmic star formation history}\label{sec:SFRDhist}
We derive the time evolution of the star formation rate density integrating, at different redshifts, the product between $\phi (L, z)$  and $\mathcal{SFR}(L)$, i.e. the SFR corresponding to a given radio luminosity $L$:
\begin{equation}
\mathcal{SFRD} (z) = \int_{L_{\rm min}}^{L_{\rm max}} \phi (L, z) \times \mathcal{SFR}(L) \,d\log L
\end{equation}
Assuming that our LF can be extrapolated at luminosities lower and higher than the actually sampled LF, we integrate over the entire luminosity range (that is, $L_{\rm min} = 0$ and $L_{\rm max} \rightarrow \infty$). The most consistent contribution to the integral comes from points located near the function knee $L_{\star}$, as the integral rapidly converges for higher luminosities. As such, the most reliable $\SFRD$ measures will be the ones where the knee is well sampled, while for the rest there will be a certain degree of uncertainty due to the fitted extrapolation \citep[this is well visible in Fig.\,8 of][]{2017A&A...602A...5N}.
\begin{figure*}
    \centering
    \includegraphics[width=1.0\textwidth]{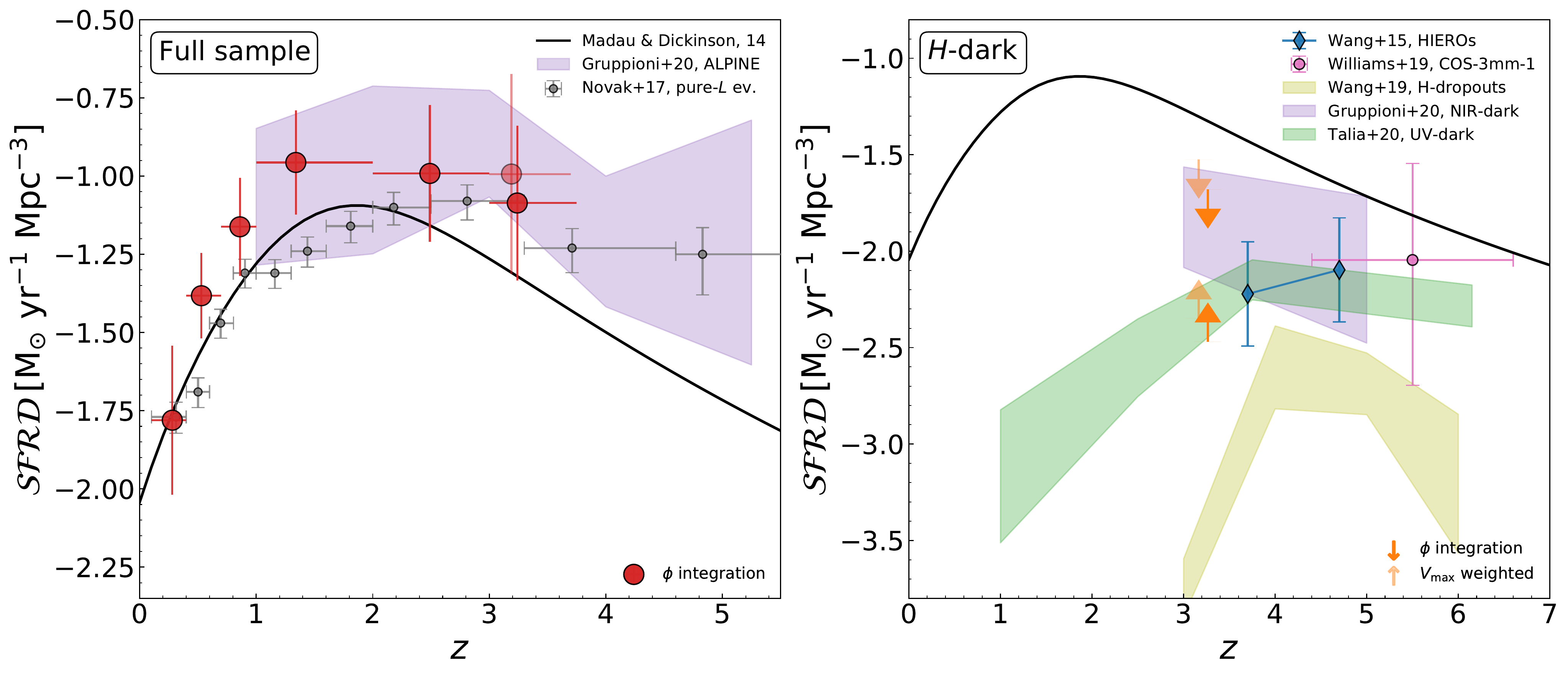}
    \caption{{\it Left panel}: the cosmic star formation rate density history for our sample of galaxies corrected per $f_{\rm AGN}$ (red points/arrows). Black curve is the best-fit function in Eq.\,15 of \citet{2014ARA&A..52..415M}, scaled to a Chabrier IMF; purple region is the $\SFRD$ measured from 56 blindly selected sources in the ALPINE survey \citep{2020A&A...643A...8G}; grey points are the pure-luminosity evolution in \citet{2017A&A...602A...5N}. {\it Right panel}: the same, limited to dark galaxies. All the values from the literature are $V_{\rm max}$-weighted. Green area is the $\SFRD$ from a sample of radio-selected UV-dark galaxies in COSMOS \citep{2021ApJ...909...23T}, yellow area is the $\SFRD$ from the 39 ALMA-detected $H$-dropouts in \citet{2019Natur.572..211W}, blue diamonds represent a sample of high-$z$ JH-blue HIEROs \citep{2016ApJ...816...84W}, magenta circle corresponds to a NIR-dark galaxy serendipitously detected in COSMOS \citep{2019ApJ...884..154W,2021arXiv210207772Z}, purple region marks the subsample of NIR-dark galaxies in the ALPINE fields \citep{2020A&A...643A...8G}. In both panels, the light red/orange points and arrows at $z > 3$ refer to the case in which we place all high-fidelity $H$-dark galaxies without a photometric redshift at $z \sim 3$.}
    \label{fig:SFRDs}
\end{figure*}
\begin{table}
    \centering
    \caption{Star formation rate density values. At $z>3$ we report also the value obtained by adding to the sample the 7 $H$-dark with high fidelity but no photo-$z$, under the assumption that they all lie at $z \sim 3$. In the last two lines we report the values for the $H$-dark galaxies alone, first considering only those with a measured photo-$z>3$, then including also the 7 ones with no photo-$z$. We report our confidence intervals, where the lower limit is derived by summing the (observed) volume-weighted SFRs, and the upper limit is derived from $\phi$.}\label{tab:SFRDs}
    \begin{tabular}{CCCCC}
    \hline
    {\rm Redshift} & {\rm From}\,\,\phi \\
    \hline
    0.1 < z < 0.4 & -1.78 \pm 0.24 \\
    0.4 < z < 0.7 & -1.38 \pm 0.14 \\
    0.7 < z < 1.0 & -1.16 \pm 0.16 \\
    1.0 < z < 2.0 & -0.96 \pm 0.17 \\
    2.0 < z < 3.0 & -0.99 \pm 0.22 \\
    z > 3.0 & -1.09 \pm 0.25 \\
    z > 3.0\,\,({\rm incl.\,\,all}\,\,H{\rm-dark}) & -0.99 \pm 0.32 \\
    \hline
    H{\rm-dark}\,\,{\rm with\,\,photo-z} & [-2.47\,;\,-1.69]\\
    {\rm All}\,\,H{\rm-dark} & [-2.35\,;\,-1.58] \\
    \hline
    \end{tabular}
\end{table}

The IR emission is directly related to the $\mathcal{SFR}$ of a source due to its nature as re-processed light coming from UV photons of newborn stars and re-emitted at longer wavelengths \citep{1998ARA&A..36..189K, 2012ARA&A..50..531K}\footnote{Rescaled for a Chabrier IMF.}
\begin{equation}
    \mathcal{SFR}\,\left[ M_{\odot}\,{\rm yr}^{-1} \right] = 2.84\times10^{-44} L_{\rm IR} \left[ {\rm erg\,\,s}^{-1} \right]
\end{equation}
As such, there is a direct link between the radio luminosity and the $\mathcal{SFR}$, via the IRRC
\begin{equation}\label{eq:SFR-L_conv}
    \mathcal{SFR}(L)\,\left[ M_{\odot}\,{\rm yr}^{-1} \right] = 10^{q_{\rm TIR} (z)} L_{1.4 {\rm GHz}} \times 10^{-24}\,\left[ W\,{\rm Hz}^{-1} \right] 
\end{equation}
with $q_{\rm TIR} (z)$ being the relation obtained in Sec.\ref{subsec:AGNs}. 

This relation holds in the redshift interval of our interest, while it may start to break down at higher redshifts ($z > 6$), when the energy densities of the Cosmic Microwave Background and the galactic magnetic field are similar, resulting in Compton cooling on the CMB from the accelerated electrons, thus leading to an underestimation of the SFR \citep{2014ApJ...784....9B}. By passing through the IRRC we are able to decompose the FIR emission in its star-formation and AGN components, and derive the SFR accordingly, something that it is not possible when directly dealing with the radio.h

The evaluated $\SFRD$ is reported in Tab.\,\ref{tab:SFRDs} and is shown in Fig.\,\ref{fig:SFRDs}, left panel, along with other estimates from the literature. At $z>3$ we give a further estimate by adding to the sample the 7 $H$-dark galaxies with a high fidelity ($>90\%$) but no photometric redshift, under the assumption that they all lie at $z \sim 3$ and have the same luminosity distribution as the rest of the sample in the same redshift bin.

Our $\phi$-inferred $\mathcal{SFRD}$ follows the well-established cosmic picture of star formation depicted in \citet{2014ARA&A..52..415M} up to redshift $\sim 1$. At earlier cosmic times we obtain systematically higher values, though compatible within the errorbars, such as the two points at $z \sim 2.51$ and $z \sim 3.25$, up to a factor $\sim 2$, resembling the (radio-inferred) evolution in \citet{2017A&A...602A...5N} and the (IR-inferred) one in \citet{2020A&A...643A...8G}. 
More recently, \citet{2022MNRAS.509.4291M} report a decreasing radio-derived SFRD at high redshift, although the deviation with respect to our and \citet{2017A&A...602A...5N}'s findings is likely a result of incompleteness due to their stellar-mass selection, which also causes them to miss out heavily obscured objects that still emit a significant amount of radio emission due to ongoing star formation, as the authors themselves point out in their conclusions.
There are not enough sources at $z > 4$ in our sample to measure the $\mathcal{SFRD}$ from the LF.

In general, our $\mathcal{SFRD}$ values are higher than the ones in \citet[][gray points in Fig\,\ref{fig:SFRDs}]{2017A&A...602A...5N}, due to a combination of cosmic variance (i.e. the point at $z \sim 0.53$) and a steeper faint end slope of the LF (1.62 vs 1.22).

\subsection{Weighting the $H$-dark contribution}\label{subsec:Hdarkweight}
We also evaluate separately the contributions of the $H$-dark galaxies to the SFRD. 
Since all but one of those with a photometric redshift are at $z \geq 3$ , we focus only on that bin.

We evaluate the LF of both the sample of $H$-dark galaxies with a measured photo-$z$, and the total sample of $H$-dark galaxies, assuming that those with no photo-$z$ lie at $z \sim 3$, similarly to what was done for the full radio sample. Then, we integrate the LF to obtain the SFRD. Such a measurement is inevitably affected by the small number of sources. We fix the turnover luminosity to the one measured with the full radio sample in the same redshift bin, and just fit the normalization $\phi_{\star}$, obtaining the $H$-dark contribution as the ratio between the two.

The assumption of an identical shape for the LF of $H$-dark galaxies and that of the full sample might bias the results. In principle, the fraction of highly obscured, $H$-dark galaxies at lower SFR is expected to be smaller than for the full population. Therefore, the $\phi$-inferred points should be considered as upper limits to the $\SFRD$. 

Following \citet{2021ApJ...909...23T} we also derive an estimate of the $\SFRD$ by summing the observed $\mathcal{SFR}$ of each galaxy, weighted by its $V_{\rm max}$, via a bootstrap analysis that takes into account the uncertainties on photometric redshift and L$_{\rm IR}$. This way we actually obtain a lower limit to the SFRD.

The results from both methods are reported in the second part of Tab.\,\ref{tab:SFRDs} and in the right panel of Fig.\,\ref{fig:SFRDs}, along with a compilation of estimates coming from literature studies of extremely obscured galaxies in the NIR. 

As we briefly mentioned in the introduction, there are multiple definitions for \emph{dark} sources across the literature. Frequently used labels (HST-, UV-, OIR-, NIR-, H-dark galaxies) refer to the fact that a common feature of these sources is that they are invisible or extremely faint at optical and/or NIR wavelengths, corresponding to the UV rest-frame as most of these galaxies are supposedly extremely dust-obscured objects at high redshift.
Such labels are obviously not absolute definitions, but they are relative to the depth of the specific band used for the selection in a given field. 
Such inhomogeneity makes it quite challenging to compare different samples, although by now various works have shown that \emph{dark} galaxies likely play an important role in the SFRD and mass assembly in the high-redshift Universe.
The comparison shown in the right panel of Fig.\,\ref{fig:SFRDs} should be read keeping this caveat in mind.

Given our derived upper and lower limits for the SFRD of $H$-dark galaxies, we find that at $z \sim 3$ they account between $\sim 3 \%$ and $\sim 25\%$ of the total star-formation activity with respect to the full radio-sample, and between $\sim 7 \%$ and $\sim 58\%$ with respect to the \citet{2014ARA&A..52..415M} estimate, which is based on UV-bright galaxies. A similar lower limit on the contribution of \emph{dark} galaxies to the total SFRD at $z\sim3$, as derived from UV-bright galaxies, is also reported by \citet{2021ApJ...909...23T} for a radio-selected sample in the COSMOS field, and by \citet{2020A&A...643A...8G}, for a sub-mm sample in the ALPINE fields.

The space density of our $H$-dark galaxies at $z > 3$ is about $1.2 \times 10^{-5}$ Mpc$^{-3}$, in fair agreement with the numbers reported in previously cited works at the same redshift, e.g. $2.0 \times 10^{-5}$ Mpc$^{-3}$ \citep{2019Natur.572..211W}, and $1.3 \times 10^{-5}$ Mpc$^{-3}$ \citep{2021ApJ...909...23T}. Existing semianalytical models \citep{2015MNRAS.451.2663H} and hydrodynamical simulations \citep{2017MNRAS.468..207S,pillepich2018} in the literature do not predict the early formation of such a large number of massive, dusty galaxies, and underestimate their number density by one to two orders of magnitude \citep[see also][]{2019Natur.572..211W}, with respect to our findings.

\section{Summary}\label{sec:Summary}

In this work we exploited a sample of radio-selected galaxies in deep 1.4GHz VLA observations of the GOODS-N field to build the radio luminosity function and measure the cosmic evolution of the star formation rate density up to $z\sim3.5$. There is a tension at redshift higher than $2$ between the steep decline in SFRD that UV-based samples  (i.e. LBGs) show and the rather flat trend derived from FIR/radio surveys. The latter are crucial to overcome the intrinsic limits of UV-based selections, i.e. poorly constrained dust correction at high-$z$, extremely dusty undetected objects.

We start from the 1.4 GHz map and catalog of \citet{2018ApJS..235...34O}, combined with ancillary multiwavelength data, and select 554 radio sources with a counterpart in the F160W catalog \citet{2019ApJS..243...22B}, plus 17 $H$-dark galaxies. We are able to assign a redshift, either spectroscopic or photometric, to all our galaxies, except for 9 $H$-dark.

We build the radio luminosity function after subtracting the AGN contamination from the radio emission. In particular, instead of removing altogether the sources with some evidence of nuclear activity from their radio excess, we estimate the fraction of AGN emission at 1.4GHz and correct the fluxes accordingly. We fit the radio LF in five redshift bins with a modified Schechter function, assuming an invariant shape for the functional form, and measure the SFRD from the integration of the LF. 

Our main finding is the evolution with redshift of the radio-inferred SFRD, which increases up to $z\sim 2$ and flattens at higher redshift. This result is consistent with other claims in the literature both from radio \citep{2017A&A...602A...5N} and sub-mm \citep{2020A&A...643A...8G} surveys, and confirms the tension at $z>3$, already reported in previous works, with respect to the SFRD estimates based on LBG samples \citep[e.g.][]{2014ARA&A..52..415M}.

We also derive the SFRD of the sub-sample of $H$-dark galaxies at $z\sim3$ and we estimate them to have a contribution to the total SFRD of $3-25\%$ and $7-58\%$ when considering the radio-based or the UV-based estimates, respectively. This result, which is consistent with other works that analyzed different samples of \emph{dark} galaxies, highlights the possibly preminent role that extremely obscured sources might play at high redshift. Dedicated follow-ups of such objects, with facilities like ALMA and the James Webb Space Telescope (JWST), would allow to robustly determine their redshift and to properly characterize their physical properties and evolution with cosmic time.

\acknowledgments
We thank the anonymous reviewer for his/her comments, that improved the work quality and flow.
We thank Frazer Owen and Lennox Cowie for kindly providing respectively the JVLA 1.4 GHz and the SCUBA-2 850 $\mu$m maps used thorough the work.
We acknowledge the support from grant PRIN MIUR 2017 - 20173ML3WW\_001.
The National Radio Astronomy Observatory is a facility of the National Science Foundation operated under cooperative agreement by Associated Universities, Inc.
This work has made use of the Rainbow Cosmological Surveys Database, which is operated by the Universidad Complutense de Madrid (UCM), partnered with the University of California Observatories at Santa Cruz (UCO/Lick,UCSC).
We acknowledge the use of Python (v 3.7) libraries in the analysis. This research made use of Photutils, an Astropy package for detection and photometry of astronomical sources (Bradley et al. 2019).
\facilities{VLA, HST, Spitzer, Herschel, CFHT, KPNO}
\software{astropy \citep{2013A&A...558A..33A}, 
        photutils \citep{larry_bradley_2019_3568287}, 
        TheTractor \citep{2016ascl.soft04008L, 2016AJ....151...36L},
        magphys \citep{2015ApJ...806..110D, 2019ApJ...882...61B, 2015ApJ...806..110D},
        sed3fit \citep{2013A&A...551A.100B},
        ultranest \citep{2014A&A...564A.125B, 2019PASP..131j8005B}}
        
\bibliography{biblio}{}
\bibliographystyle{aasjournal}

\appendix
\section{SEDs of $H$-dark galaxies}\label{app:Hdark}
\begin{figure*}[b!]
    \centering
    \includegraphics[width=0.85\textwidth]{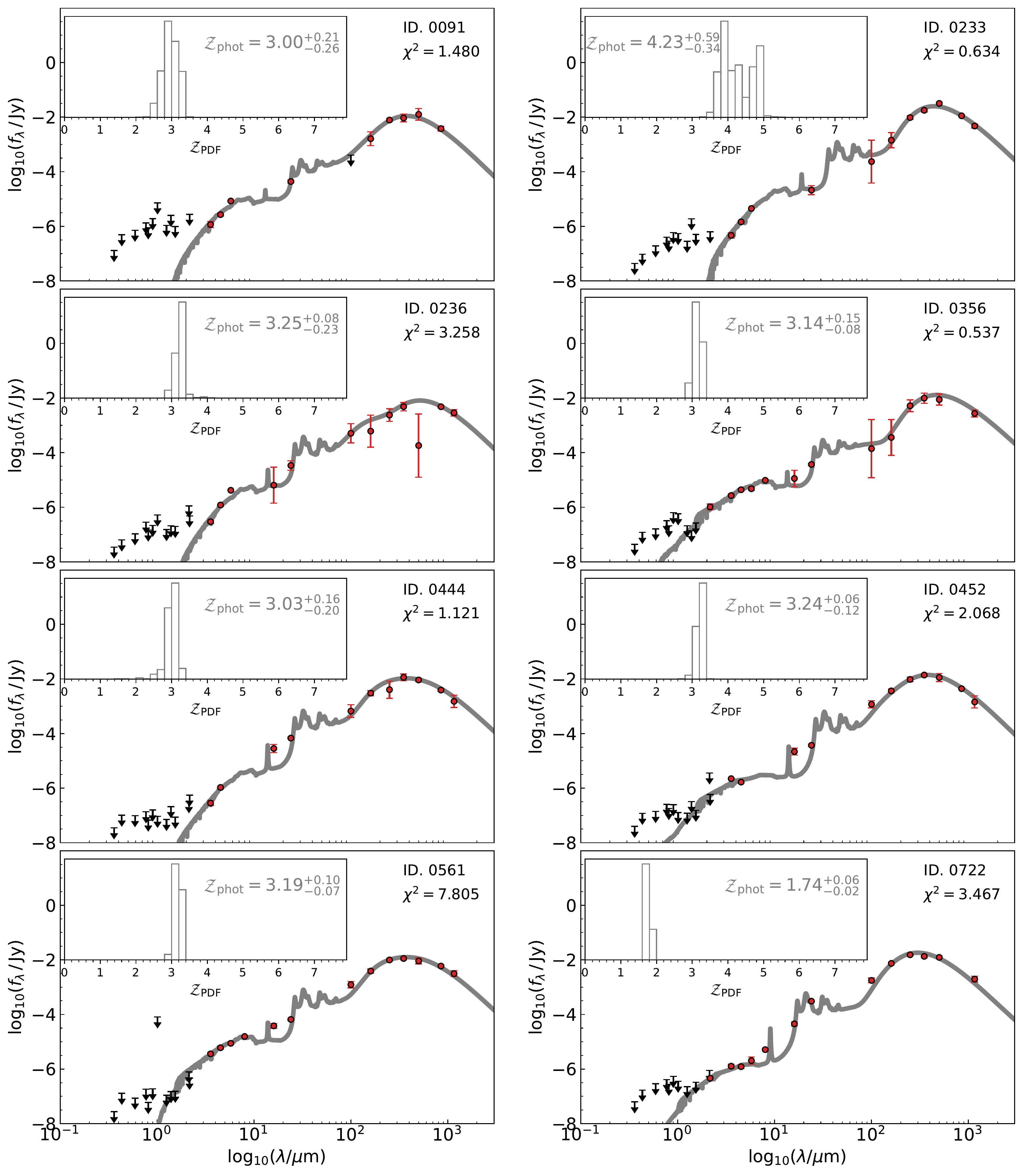}
    \caption{Best-fit SEDs (gray) of the $H$-dark sources in our radio-selected sample; red circles indicate the photometric measurements, while black arrows stand for upper limits. The redshift probability distribution function is also shown in the inset. We also reported the reduced-$\chi^2$.}
    \label{fig:SEDs}
\end{figure*}

In this Appendix we show the photometric data plus best-fit SEDs of the 8 $H$-dark galaxies in our sample for which we could derive a photometric redshift (Fig.\,\ref{fig:SEDs}).

\end{document}